\def\kB{k_{\text{B}}}
\def\epsilonF{\epsilon_{\text{F}}}
\def\kF{k_{\text{F}}}

\def\NF{N_{\text{F}}}
\def\me{m_{\text{e}}}

\def\Tc{T_{\text{c}}}
\def\pc{p_{\,\text{c}}}
\def\Hcone{{H_{\text{c1}}}}
\def\Hctwo{{H_{\text{c2}}}}
\def\gso{g_{\text{so}}}
\def\be{\begin{equation}}
\def\ee{\end{equation}}
\def\bea{\begin{eqnarray}}
\def\eea{\end{eqnarray}}
\def\bse{\begin{subequations}}
\def\ese{\end{subequations}}

\documentclass[prb,twocolumn,showpacs,amsmath,amssymb,eqsecnum,preprintnumbers]
                                                                        {revtex4}
\usepackage{graphicx}
\usepackage{dcolumn}
\usepackage{bm}
\usepackage{rotating} 
\begin{document}
\title{Ordered Phases of Itinerant Dzyaloshinsky-Moriya Magnets and Their Electronic
       Properties}
\author{Kwan-yuet Ho and T.R. Kirkpatrick}
\affiliation{Institute for Physical Science and Technology,
             and Department of Physics, University of Maryland, College Park,
             MD 20742, USA
             }
\author{Yan Sang and D. Belitz}
\affiliation{Department of Physics, and Institute of Theoretical Science,
             University of Oregon, Eugene, OR 97403, USA
             }
\date{\today}

\begin{abstract}
A field theory appropriate for magnets that display helical order due to the
Dzyaloshinsky-Moriya mechanism, a class that includes MnSi and FeGe, is used to
derive the phase diagram in a mean-field approximation. The helical phase, the
conical phase in an external magnetic field, and recent proposals for the
structure of the A-phase and the non-Fermi-liquid region in the paramagnetic
phase are discussed. It is shown that the orientation of the helical pitch
vector along an external magnetic field within the conical phase occurs via two
distinct phase transitions. The Goldstone modes that result from the long-range
order in the various phases are determined, and their consequences for
electronic properties, in particular the specific heat, the single-particle
relaxation time, and the electrical and thermal conductivities, are derived.
Various aspects of the ferromagnetic limit, and qualitative differences between
the transport properties of helimagnets and ferromagnets, are also discussed.
\end{abstract}

 \maketitle

\tableofcontents

\section{Introduction}
\label{sec:I}

\subsection{Dzyaloshinsky-Moriya magnets}
\label{subsec:I.A}

Helical magnets are systems in which long-range magnetic order takes the form
of a helix or spiral, such that in any given plane perpendicular to a preferred
direction there is ferromagnetic order, but the direction of the magnetization
rotates as one goes along the preferred axis. The pitch vector ${\bm q}$ of the
helix points in the preferred direction, and its modulus $q \equiv \vert{\bm
q}\vert$ is the pitch wave number, with $2\pi/q$ equal to the helical
wavelength. One mechanism for stabilizing this type of order over the
homogeneous ferromagnet was pointed out by
Dzyaloshinksky\cite{Dzyaloshinsky_1958} and Moriya.\cite{Moriya_1960} It relies
on the spin-orbit interaction, which can lead, for certain lattice structures,
to a term of the form ${\bm M}\cdot({\bm\nabla}\times{\bm M})$ in the
Hamiltonian or action, with ${\bm M}$ the magnetic order parameter. The
presence of such a chiral term implies that a homogeneously magnetized state
can always gain energy by means of a nonzero curl of the magnetization, which
leads to a helical ground state. In a rotationally invariant system the
direction of the pitch vector would be arbitrary (analogous to the arbitrary
direction of the magnetization in an isotropic ferromagnet). However, the
underlying lattice leads, via crystal-field effects, to a pinning of the helix
in certain directions determined by the lattice. The crystal-field effects are
due to the spin-orbit interaction, as is the Dzyaloshinsky-Moriya (DM)
mechanism for helical order itself. Since the spin-orbit interaction is weak on
an atomic or microscopic scale, with a dimensionless coupling constant $\gso
\ll 1$, this leads to a hierarchy of energy or length scales that can be
classified according to their dependence on powers of $\gso$. Finally, an
external magnetic field provides an additional energy scale, and couples to the
helix via an induced homogeneous component of the magnetization. This leads to
a rich phase diagram that is the topic of the present paper.

\subsubsection{The phase diagram of MnSi}
\label{subsubsec:I.A.1}

\begin{figure}[t]
\vskip -0mm
\includegraphics[width=6.0cm]{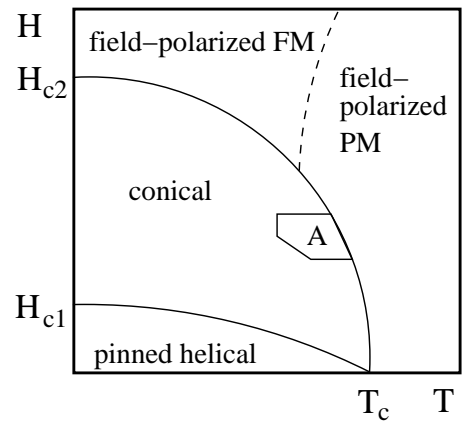}
\caption{Schematic phase diagram of MnSi in the $H$-$T$ plane showing the
 helical, conical, and A-phases, as well as the field-polarized ferromagnetic
 (FM) and paramagnetic (PM) states. See the text for a description of these
 phases.}
\label{fig:1}
\end{figure}

A very well-studied helimagnet of DM-type is the metallic system MnSi, which we
will concentrate on; a very similar, but less extensively studied material is
FeGe.\cite{Lebech_Bernhard_Freltoft_1989} In zero magnetic field, and at
ambient pressure, MnSi displays long-ranged helical order with $2\pi/q \approx
180\,\AA$ below a temperature $\Tc \approx 30\,{\text
K}$.\cite{Ishikawa_et_al_1976} MnSi crystallizes in the cubic B20 structure,
and the helix is observed to be pinned in the $\langle
1,1,1\rangle$-directions.\cite{Lundgren_et_al_1970, Ishikawa_et_al_1977} The
corresponding space group is P2$_1$3.\cite{Nakanishi_et_al_1980} In a magnetic
field $H$, the helix is superimposed by a homogeneous component of the
magnetization, which leads to the so-called conical
phase.\cite{Ishikawa_et_al_1977} A magnetic field in the $[0,0,1]$-direction
tilts the helix away from $\langle 1,1,1\rangle$ until the ${\bm q}$-vector
aligns with the magnetic field at a critical field strength $\Hcone$. With
increasing $H$ the amplitude of the helix decreases, and finally vanishes at a
critical field $\Hctwo$, where the system enters a field-polarized
ferromagnetic phase. Inside the conical phase at intermediate fields near $\Tc$
there is a region known as the A-phase.\cite{Ishikawa_Arai_1984} This was
thought to represent a helix whose pitch vector is perpendicular to the
magnetic field,\cite{Lebech_1993, Grigoriev_et_al_2006a} but recently has been
interpreted as a topological phase where three helices with co-planar ${\bm
q}$-vectors form a skyrmion-like structure.\cite{Muhlbauer_et_al_2009} The
schematic phase diagram in the $H$-$T$-plane as observed experimentally is
shown in Fig.\ \ref{fig:1}. In addition to the phases shown, a possible second
transition and spin-liquid phase at $H=0$ just above $\Tc$ has been reported
recently.\cite{Pappas_et_al_2009}

\begin{figure}[t]
\vskip -0mm
\includegraphics[width=7.0cm]{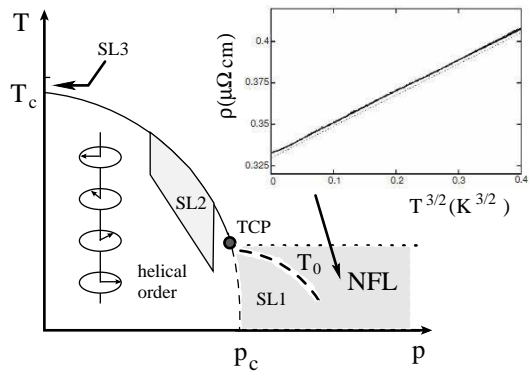}
\caption{Schematic phase diagram of MnSi in the $T$-$p$ plane. The tricritical
 point (TCP) separates a line of second order transitions (solid line) from a
 line of first-order transitions (dashed line). The inset shows the resistivity
 data from Ref.\ \onlinecite{Pfleiderer_Julian_Lonzarich_2001} in the
 non-Fermi-liquid (NFL) region. The boundary of the NFL region (dotted line)
 is not sharp. SL1, SL2, and SL3 refer to the possible spin-liquid phases or
 regions reported in Refs.\ \onlinecite{Pfleiderer_et_al_2004} and
 \onlinecite{Uemura_et_al_2007}, \onlinecite{Lee_et_al_2009}, and
 \onlinecite{Pappas_et_al_2009}, respectively.}
\label{fig:2}
\end{figure}

\begin{figure}[t]
\vskip -0mm
\includegraphics[width=8.0cm]{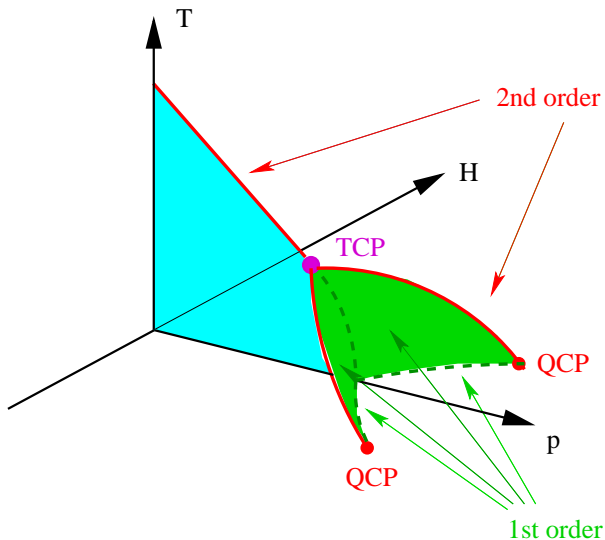}
\caption{Schematic phase diagram of MnSi in the space spanned by $T$, $p$, and
 $H$ showing the tricritical wings and quantum critical points (QCP).}
\label{fig:3}
\end{figure}

Another interesting aspect of MnSi is its sensitivity to hydrostatic pressure.
With increasing pressure $p$, the magnetic transition temperature $\Tc$
decreases until it vanishes at $p = p_{\,\text{c}} \approx
15\,\text{kbar}$.\cite{Pfleiderer_et_al_1997} The transition is second order or
very weakly first order above a temperature of approximately $10\,{\text{K}}$,
and strongly first order at lower temperatures, with a tricritical point
separating the two regimes. These features have been explained as universal
properties of quantum ferromagnets in an approximation that neglects the
helical order at longer length scales.\cite{Belitz_Kirkpatrick_Vojta_1999} In
the paramagnetic region at pressures up to approximately $2p_{\,\text{c}}$ and
at temperatures $T\alt 10\,\text{K}$, strong non-Fermi-liquid transport
properties are observed, with the electrical resistivity $\rho_{\text{el}}$
displaying a temperature dependence $\rho_{\text{el}} \propto T^{3/2}$ over
almost three decades in temperature.\cite{Pfleiderer_Julian_Lonzarich_2001} The
origin of this behavior has recently been proposed to be a combination of
columnar spin textures and very weak quenched
disorder.\cite{Kirkpatrick_Belitz_2010} In a smaller region, above
$p_{\,\text{c}}$ and below a pressure-dependent temperature $T_0$, short-range
(spin-liquid) helical order has been observed\cite{Pfleiderer_et_al_2004,
Uemura_et_al_2007} and various explanations in terms of analogs of blue phases
in liquid crystals have been proposed.\cite{Tewari_Belitz_Kirkpatrick_2006,
Binz_Vishvanath_Aji_2006, Rossler_Bogdanov_Pfleiderer_2006,
Fischer_Shah_Rosch_2008} Recently, an anomalous Hall effect has been observed
at intermediate pressures and below $\Tc$ in the helically ordered phase, and
it has been suggested that this feature is related to the short-ranged order
observed above $p_{\,\text{c}}$.\cite{Lee_et_al_2009} The phase diagram in the
$T$-$p$ plane is shown in Fig.\ \ref{fig:2}.
%
If a magnetic field is applied in the vicinity of $\pc$, tricritical wings,
i.e., surfaces of first-order transitions, emerge from the tricritical point
that are believed to end in a pair of quantum critical points in the $T=0$
plane.\cite{Pfleiderer_Julian_Lonzarich_2001} This feature, which is depicted
in Fig.\ \ref{fig:3}, has been explained theoretically in Ref.\
\onlinecite{Belitz_Kirkpatrick_Rollbuehler_2005}.

\subsubsection{Properties of the helical phase}
\label{subsubsec:I.A.2}

If one ignores crystal-field effects, i.e., in an isotropic model, the
long-range helical order in the helical phase ($H=0$, $T < \Tc$, $p < \pc$)
breaks the translational symmetry. This spontaneous breaking of a continuous
symmetry leads to a Goldstone mode, the
helimagnon.\cite{Belitz_Kirkpatrick_Rosch_2006a} The dispersion relation is
anisotropic, and given by
\be
\omega_{\,\text{HM}}({\bm k}) = \sqrt{c_{||}k_{||}^2 + c_{\perp}{\bm
k}_{\perp}^4}.
\label{eq:1.1}
\ee
Here the wave vector ${\bm k} = (k_{||},{\bm k}_{\perp})$ has been decomposed
into components parallel and perpendicular, respectively, to the helical pitch
vector ${\bm q}$, and $c_{||}$ and $c_{\perp}$ are elastic constants. The
absence of a term proportional to ${\bm k}_{\perp}^2$ under the square root in
Eq.\ (\ref{eq:1.1}) reflects rotational symmetry and is analogous to the nature
of the Goldstone mode in either smectic or cholesteric liquid
crystals.\cite{Chaikin_Lubensky_1995} Crystal-field effects that couple the
electrons to the underlying lattice break the rotational invariance and lead to
a ${\bm k}_{\perp}^2$ term with a small prefactor, i.e., the dispersion
relation becomes phonon-like. Both the single-particle relaxation rate $1/\tau$
that results from the scattering of electronic quasi-particles by helimagnons
and the thermal resistivity $\rho_{\text{th}}$ have a non-Fermi-liquid
temperature dependence $1/\tau \propto \rho_{\text{th}} \propto T^{3/2}$. In
contrast, the transport relaxation rate that determines the electrical
conductivity $\rho_{\text{el}}$ in a Boltzmann approximation has a temperature
dependence $1/\tau^{\text{tr}}_{\text{el}} \propto \rho_{\text{el}} \propto
T^{5/2}$.\cite{Belitz_Kirkpatrick_Rosch_2006b} The properties of the helical
phase in the presence of weak quenched disorder have also been
investigated.\cite{Kirkpatrick_Belitz_Saha_2008a,
Kirkpatrick_Belitz_Saha_2008b}

In the present paper we will discuss the derivation of the phase diagram
sketched above, as well as some refinements, within the context of a mean-field
theory and with an emphasis on the $H$-$T$ plane. We will also determine the
nature of the Goldstone modes, and their consequences for electronic
properties, in various observed or proposed phases with long-range order. In
addition to the specific heat, the single-particle relaxation rate, and the
electrical conductivity we will consider the thermal resistivity
$\rho_{\text{th}}$, whose temperature dependence is the same as that of the
single-particle relaxation rate.

\subsection{Model, and energy scales}
\label{subsec:I.B}

\subsubsection{LGW functional}
\label{subsubsec:I.B.1}

We will consider a Landau-Ginzburg-Wilson (LGW) functional for a
three-dimensional order parameter (OP) field ${\bm M} = (M_1,M_2,M_3)$ whose
expectation value is proportional to the magnetization. We will consider an
action that is appropriate for the helical magnets MnSi and FeGe, which
crystallize in the cubic B20 structure with space group P2$_1$3. We will
organize the action according to the dependence of its various constituents on
powers of the spin-orbit interaction $\gso$. Within this scheme, we write
\be
{\cal A} = {\cal A}_{\text{H}} + {\cal A}_{\text{DM}} + {\cal A}_{\text{cf}},
\label{eq:1.2}
\ee
where
\bea
{\cal A}_{\text{H}} &=& \int_V d{\bm x}\ \left[\frac{t}{2}\,{\bm M}^2({\bm x})
+ \frac{a}{2}\,\left(\nabla{\bm M}({\bm x})\right)^2 \right.
\nonumber\\
&&\hskip 25pt + \frac{d}{2}\,\left({\bm\nabla}\cdot{\bm M}({\bm x})\right)^2 +
\frac{u}{4}\left({\bm M}^2({\bm x})\right)^2
\nonumber\\
&&\hskip 25pt + \frac{w}{4}\left({\bm\nabla}{\bm M}^2({\bm x})\right)^2 - {\bm
H}\cdot{\bm M}({\bm x})\biggr]
\label{eq:1.3}
\eea
is the action for an isotropic classical Heisenberg ferromagnet in a
homogeneous external magnetic field ${\bm H}$. $\int_V d{\bm x}$ denotes a
real-space integral over the system volume. $(\nabla{\bm M})^2$ stands for
$\sum_{i,j=1}^3 \partial_i\,M_j\,\partial^i\,M^j$, with $\partial_i \equiv
\partial/\partial x_i$ the components of the gradient operator ${\bm\nabla}
\equiv (\partial_1,\partial_2,\partial_3) \equiv (\partial_x, \partial_y,
\partial_z)$. $t$, $a$, $d$, $u$, and $w$ are the parameters of the Landau
theory and are related to the microscopic energy and length scales (see Sec.\
\ref{subsubsec:I.B.2} below); they are of zeroth order in the spin-orbit
coupling $\gso$. Equation (\ref{eq:1.3}) contains all analytic terms invariant
under co-rotations of real space and OP space up to quartic order in ${\bm M}$
and bi-quadratic order in ${\bm M}$ and ${\bm\nabla}$.\cite{dipole_footnote} We
have added one higher-order term, with coupling constant $w$, as an example of
a class of terms that can stabilize unusual phases in helimagnets, although
they are not of qualitative importance in ferromagnets. We will consider this
term only in Appendix \ref{app:A}.

\bse
\label{eqs:1.4}
\be
{\cal A}_{\text{DM}} = \frac{c}{2}\int_V d{\bm x}\ {\bm M}({\bm
x})\cdot({\bm\nabla}\times{\bm M}({\bm x}))
\label{eq:1.4a}
\ee
is the chiral Dzyaloshinksy-Moriya term that favors a nonvanishing curl of the
magnetization. The existence of this term hinges on the spin-orbit coupling, as
well as on the system not being invariant with respect to spatial inversion
(due to the linear dependence on the gradient operator). The coupling constant
$c$ is linear in $\gso$, and on dimensional grounds we have
\be
c = a\,\kF\,\gso,
\label{eq:1.4b}
\ee
\ese
with $\kF$ the Fermi wave number which serves as the microscopic inverse length
scale. In this context, this can be considered the definition of $\gso$.

The preceding contributions to the action are all invariant under either
separate rotations, or co-rotations, in spin (or magnetization) space and real
space. The spin-orbit interaction couples the electron spin, and hence the
magnetization, to the underlying lattice. Therefore, in addition to the
rotationally invariant terms, any term that is invariant under elements of the
space group connected with the crystal lattice is allowed. For the B20
structure of MnSi and FeGe, the appropriate space group is P2$_1$3. To quartic
order in ${\bm M}$, and bi-quadratic order in ${\bm\nabla}$ and ${\bm M}$, the
allowed terms in the action are the crystal-field terms
\bse
\label{eqs:1.5}
\bea
{\cal A}_{\text{cf}} &=& \int_V d{\bm x}\ \sum_{i=1}^3 \left[\frac{b}{2}
(\partial_i\,M_i({\bm x}))^2 + \frac{b_1}{2} (\partial_i\,M_{i+1}({\bm x}))^2
\right.
\nonumber\\
&& \left. \hskip 90pt + \frac{v}{4}\,M_i^4({\bm x})\right]
\label{eq:1.5a}
\eea
where $M_4 \equiv M_1$. The last term is the usual cubic anisotropy that is
always present in a magnet on a cubic lattice, and
\be
v = u' \gso^4
\label{eq:1.5b}
\ee
with $\vert u'\vert \approx u$. Of the gradient-squared terms, the first one
also has cubic symmetry; the second one does not, but is invariant under
elements of P2$_1$3. On dimensional grounds, we have
\bea
b &=& a' \gso^2,
\nonumber\\
b_1 &=& a_1'\, \gso^2,
\label{eq:1.5c}
\eea
\ese
with $\vert a' \vert \approx \vert a_1' \vert \approx a$.

\subsubsection{Length and energy scales}
\label{subsubsec:I.B.2}

The various contributions to the action ${\cal A}$, and their dependencies on
$\gso$, imply a hierarchy of energy scales and corresponding wave number or
length scales. At zeroth order in $\gso$, we have the microscopic scale, which
is represented by the Fermi energy $\epsilonF$ and the Fermi wave number $\kF$.
Fluctuations renormalize this to the critical scale, which is represented by
the magnetic ordering temperature $\Tc$ and the corresponding length scale. The
physics at these scales is described by ${\cal A}_{\text{H}}$, Eq.\
(\ref{eq:1.3}).

The chiral DM term is balanced by the rotationally invariant gradient squared
term in Eq.\ (\ref{eq:1.3}) that makes magnetization gradients energetically
costly. As a result, the relevant gradient or momentum scale is of $O(\gso)$,
and hence the chiral wave number scale is given by the microscopic scale times
$\gso$. This determines the parameters of the helix, in particular the helical
pitch wave number $q \propto \gso$. In MnSi and FeGe, this wave number scale is
on the order of 100 times smaller than the microscopic scale. ${\cal
A}_{\text{DM}}$ contains one explicit factor of $\gso$ and one gradient, and
hence its contribution to the free energy is of $O(\gso^2)$. The physics at
this scale is described by ${\cal A}_{\text{DM}}$ in conjunction with ${\cal
A}_{\text{H}}$.

At fourth order in $\gso$, crystal-field effects come into play. They pin the
helix, are small compared to the chiral energy scale by another factor of
$\gso^2$, and are described by ${\cal A}_{\text{cf}}$, Eqs.\ (\ref{eqs:1.5}).
Since gradients are effectively of $O(\gso)$, see above, the contributions of
all three terms in ${\cal A}_{\text{cf}}$ to the free energy are of
$O(\gso^4)$.

Finally, the external magnetic field sets a scale that is continuously tunable.

\section{Phase Diagram}
\label{sec:II}

We now derive the mean-field phase diagram for systems described by the action
given in Eqs.\ (\ref{eq:1.2}) - (\ref{eqs:1.5}). We will use the hierarchy of
energy scales explained in Sec.\ \ref{subsec:I.B} to show how a more and more
sophisticated phase diagram emerges as one keeps effects of higher and higher
order in $\gso$.

To do so, we consider field configurations of the following form:
\bse
\label{eqs:2.1}
\be
{\bm M}({\bm x}) = {\bm m}_0 + m_1^+\,{\bm{\hat e}}_+ \cos({\bm q}\cdot{\bm x})
+ m_1^-\,{\bm{\hat e}}_-\sin({\bm q}\cdot{\bm x}).
\label{eq:2.1a}
\ee
Here ${\bm m}_0$ is a homogeneous component of the magnetization, $m_1^{\pm}$
are amplitudes of Fourier components with wave vector ${\bm q}$, and ${\bm{\hat
e}}_{\pm}$ are two unit vectors that form a right-handed {\it dreibein}
together with ${\bm q}$:
\be
{\bm{\hat e}}_+ \times {\bm{\hat e}}_- = {\bm{\hat q}} \quad,\quad {\bm{\hat
q}} \times {\bm{\hat e}}_+ = {\bm{\hat e}}_- \quad,\quad {\bm{\hat e}}_- \times
{\bm{\hat q}} = {\bm{\hat e}}_+ \ ,
\label{eq:2.1b}
\ee
\ese
where ${\bm{\hat q}} = {\bm q}/q$. The sinusoidal terms in Eq.\ (\ref{eq:2.1a})
describe a helix with pitch vector ${\bm q}$. The helix is in general
elliptically polarized, and it is useful to define a polarization parameter
\be
\pi = m_1^-/m_1^+,
\label{eq:2.2}
\ee
Special cases are circular polarization, $\pi = 1$, and linear polarization,
$\pi = 0$ or $\pi = \infty$. The motivation for the {\it ansatz} Eqs.\
(\ref{eqs:2.1}) is provided by the fact that it gives the functional form of
the global minimum of the action ${\cal A}$, Eqs.\ (\ref{eq:1.2}) -
(\ref{eqs:1.5}), if one neglects the crystal-field terms ${\cal
A}_{\text{cf}}$; that is, for the action up to $O(\gso^2)$, see Ref.\
\onlinecite{Muhlbauer_et_al_2009} and Sec.\ \ref{subsec:II.B} below.

\subsection{$O(\gso^0)$: Ferromagnet}
\label{subsec:II.A}

To zeroth order in $\gso$ the system is approximated by a ferromagnet.
According to the action ${\cal A}_{\text{H}}$, Eq.\ (\ref{eq:1.3}), for $H=0$
there is a second-order phase transition which in mean-field approximation
occurs at $t=0$. For $H\neq 0$ there is a crossover at $t=0$ from a
field-polarized paramagnetic state, where the magnetization extrapolates to
zero for $H\to 0$, to a field-polarized ferromagnetic state where the
magnetization extrapolates to ${\bm m}_0 = \sqrt{-t/u}\,{\hat{\bm H}}$. The
free energy density in mean-field approximation, $f = {\cal A}/V$, in a zero
field is
\be
f = -t^2/4u.
\label{eq:2.3}
\ee
In a nonzero field, one has
\bse
\label{eqs:2.4}
\be
f = \frac{t}{2}\,m_0^2 + \frac{u}{4}\,m_0^4 - Hm_0,
\label{eq:2.4a}
\ee
where $m_0$ is the solution of the mean-field equation of state
\be
t\,m_0 + u\,m_0^3 = H.
\label{eq:2.4b}
\ee
\ese
This is just a classical Heisenberg model, so it cannot explain the tricritical
point and the associated tricritical wings. The latter emerge within a
renormalized mean-field theory that takes into account the coupling of the
magnetization to other electronic degrees of
freedom.\cite{Belitz_Kirkpatrick_Vojta_1999} This leads to a
fluctuation-induced first-order transition in analogy to the case of the
nematic-to-smectic-A transition in liquid
crystals,\cite{Belitz_Kirkpatrick_Vojta_2005} as well as to the tricritical
wings and the quantum-critical points in an external magnetic
field.\cite{Belitz_Kirkpatrick_Rollbuehler_2005} The ferromagnetic
approximation suffices for understanding the gross features of the phase
diagram in the $T$-$p$ plane.

\subsection{$O(\gso^2)$: Helimagnet, conical phase}
\label{subsec:II.B}

To second order in $\gso$ we need to add the DM term, Eq.\ (\ref{eq:1.4a}), to
the action. It is obvious that this term favors a nonzero curl of the
magnetization, with the direction of the curl depending on the sign of $c$.
However, the spatial variation of ${\bm M}$ will be limited by the other
gradient terms in the action, the $(\nabla{\bm M})^2$ term in particular. We
thus expect a spatial modulation of ${\bm M}$ on a length scale on the order of
$a/c$. It is easy to check that the ansatz, Eqs.\ (\ref{eqs:2.1}), with $\pi =
1$, i.e., $m_1^- = m_1^+ \equiv m_1$, and
\bse
\label{eqs:2.5}
\bea
{\bm q} &=& q\,{\hat{\bm H}},
\label{eq:2.5a}\\
{\bm m}_0 &=& m_0\,{\hat{\bm H}},
\label{eq:2.5b}\\
m_0 &=& H/(cq - aq^2),
\label{eq:2.5c}\\
m_1^2 &=& \frac{1}{u}\,\left(-t + cq -aq^2 - \frac{H^2}{(cq - aq^2)^2}\right),
\label{eq:2.5d}
\eea
\ese
solves the saddle-point equations for the action ${\cal A}_{\text{H}} + {\cal
A}_{\text{DM}}$. In order to determine $q$, we extremize the resulting free
energy with respect to $q$. $q = c/2a$ is a solution for all values of $H$.
Finally, we need to ascertain that the solution is a minimum, which turns out
to be the case for $t < aq^2$ and $H < aq^2 \sqrt{-(t-aq^2)/u}$. We thus find
that the field configuration
\bse
\label{eqs:2.6}
\be
{\bm M}({\bm x}) = m_0\,{\hat{\bm H}} + m_1\,\left[{\hat{\bm
e}}_+\,\cos(q{\hat{\bm H}}\cdot{\bm x}) + {\hat{\bm e}}_-\,\sin(q{\hat{\bm
H}}\cdot{\hat{\bm x}})\right],
\label{eq:2.6a}
\ee
with ${\hat{\bm e}}_1$, ${\hat{\bm e}}_2$, ${\hat{\bm H}}$ forming a {\it
dreibein}, and
\bea
q &=& c/2a,
\label{eq:2.6b}\\
m_0 &=& H/aq^2,
\label{eq:2.6c}\\
m_1 &=& \sqrt{-r/u}\,\sqrt{1 - (H/\Hctwo)^2},
\label{eq:2.6d}
\eea
\ese
where
\bse
\label{eqs:2.7}
\bea
r &=& t - aq^2,
\label{eq:2.7a}\\
\Hctwo &=& aq^2\sqrt{-r/u},
\label{eq:2.7b}
\eea
\ese
minimizes the free energy in the parameter range $r < 0$, $H < \Hctwo$. We will
refer to this state as the aligned conical state (ACS), to distinguish it from
the perpendicular conical state discussed in Appendix \ref{app:A}. The ACS is
actually a global minimum, as can be seen by writing the action as a sum of
positive semi-definite terms that are individually minimized by this
state.\cite{Muhlbauer_et_al_2009} The mean-field free energy density in that
range is
\be
f = \frac{-1}{2}\,\left(\frac{r^2}{2u} + \frac{H^2}{aq^2}\right).
\label{eq:2.8}
\ee
Equations (\ref{eqs:2.6}) describe the helical phase for $H=0$ and the conical
phase for $0 < H < \Hctwo$. Comparing Eqs.\ (\ref{eq:2.3}) and (\ref{eq:2.8})
we see that the helical transition preempts the ferromagnetic one.

For $H \to \Hctwo$ from below the helical component of the magnetization
vanishes, and the free energy, Eq.\ (\ref{eq:2.8}), approaches that of the
ferromagnet, Eqs.\ (\ref{eqs:2.4}). For $H > \Hctwo$ the equation of state and
the free energy for the DM action ${\cal A}_{\text{H}} + {\cal A}_{\text{DM}}$
are the same as for a ferromagnet with action ${\cal A}_{\text{H}}$.

These considerations account for the structure of the phase diagram shown in
Fig.\ \ref{fig:1} except for the field $\Hcone$ and the A-phase.

\subsection{$O(\gso^2)$: Helimagnet, A-phase}
\label{subsec:II.C}

Recent neutron-scattering experiments by M{\"u}hlbauer et
al.\cite{Muhlbauer_et_al_2009} are consistent with the notion that the A-phase
is characterized by spin textures that form line defects in the direction of
the magnetic field, with the lines forming a hexagonal lattice, the A-crystal,
in the plane perpendicular to the field. This experimental observation led the
authors of Ref.\ \onlinecite{Muhlbauer_et_al_2009} to suggest a skyrmion state
consisting of three co-planar helices as underlying the A-phase:
\bea
{\bm M}({\bm x}) &=& m_0 {\hat{\bm H}}
\nonumber\\
&&\hskip -60pt + m_1 \sum_{i=1}^{3} \left[{\hat{\bm e}}^{(i)}_+\, \cos({\bm
q}^{(i)}\cdot{\bm x}) + {\hat{\bm e}}^{(i)}_-\, \sin({\bm q}^{(i)}\cdot{\bm
x})\right],
\label{eq:2.9}
\eea
where the pitch vectors ${\bm q}^{(i)}$ all have the same modulus $q$, are
perpendicular to ${\bm H}$, the angle between adjacent pitch vectors is
$2\pi/3$, and ${\hat{\bm e}}^{(i)}_+$, ${\hat{\bm e}}^{(i)}_-$, and ${\bm
q}^{(i)}$ form a right-handed {\it dreibein} for each value of $i$. The values
of $q$, $m_0$, and $m_1$ are obtained by minimizing the action.

\begin{figure}[t]
\vskip -0mm
\includegraphics[width=8.0cm]{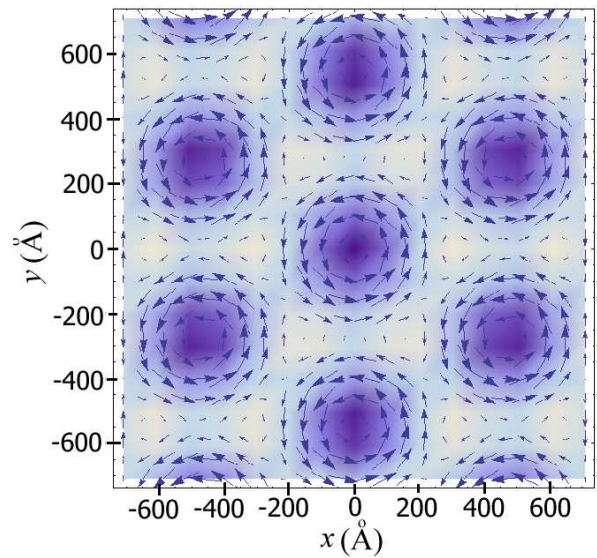}
\caption{Spin configuration as given by Eq.\ (\ref{eq:2.9}) with the ${\bm q}$
 vectors in the $x$-$y$ plane, $q = 0.0133 \AA^{-1}$, $m_0 = 0.0146$, and
 $m_1 = 0.0323$. The arrows represent projections of the spins into the plane.
 The dark and light regions denote spin directions antiparallel and parallel to
 ${\hat z}$, respectively.}
\label{fig:4}
\end{figure}

The free energy difference between this state and the conical one has a minimum
at $H \approx 0.4\,\Hctwo$, but it is still positive even at the minimum.
However, Ref.\ \onlinecite{Muhlbauer_et_al_2009} found that taking into account
Gaussian fluctuations stabilizes the state with respect to the conical one. It
should be noted that Eq.\ (\ref{eq:2.9}) is not a solution of the saddle-point
equations for the action given by Eqs.\ (\ref{eq:1.2})- (\ref{eqs:1.5}), and
therefore cannot be a true local minimum of the free energy. Also, the relation
between this {\it ansatz} and what are commonly called skyrmionic spin
configurations, which {\em are} solutions of the saddle-point equations (see
Ref.\ \onlinecite{Rossler_Bogdanov_Pfleiderer_2006} and references therein), is
not clear. Most likely it represents a single-Fourier-component approximation
to a saddle point and is a Bloch-type description of skyrmions as opposed to a
Wannier-type description of isolated skyrmions in Ref.\
\onlinecite{Rossler_Bogdanov_Pfleiderer_2006}. In any case, it describes a
hexagonal array of line defects with the spin antiparallel to the magnetic
field at the defect centers, and parallel at points on the cell boundaries, see
Fig.\ \ref{fig:4}, in qualitative agreement with the neutron scattering data.

Earlier, Grigoriev et al.\cite{Grigoriev_et_al_2006a} had proposed a
single-helix state with the pitch vector oriented perpendicular to the external
field. Although current experimental evidence favors a skyrmion state as a
candidate for the A-phase, it is still of interest to discuss such a
perpendicular conical state (PCS), since it might be a viable candidate for the
ground state in some other part of the phase diagram of helimagnets. We
therefore briefly discuss the PCS and its properties in Appendix \ref{app:A}.

\subsection{$O(\gso^2)$: Helimagnet, NFL region}
\label{subsec:II.D}

\begin{figure}[t]
\vskip -0mm
\includegraphics[width=5.0cm]{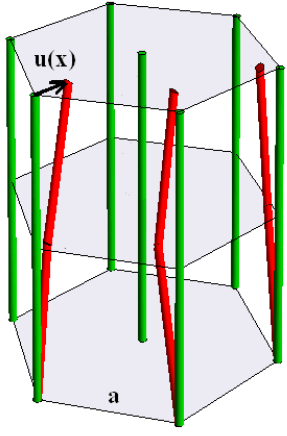}
\caption{Schematic rendering of the hexagonal lattice of skyrmion lines shown
 in Fig.\ \ref{fig:4} and of fluctuations about this state. $a$ is the lattice
 constant, and ${\bm u}({\bm x})$ is a displacement vector.}
\label{fig:5}
\end{figure}

The skyrmion lattice approximately described by Eq.\ (\ref{eq:2.9}) can melt,
which will lead to a skyrmion liquid. In such a state the line defects shown in
Fig.\ \ref{fig:4} still exist, but they no longer form a lattice. Rather, their
fluctuations in the plane perpendicular to the line, which are illustrated in
Fig.\ \ref{fig:5}, have become so large that the long-range order is destroyed.
Such a state has recently been proposed to represent the NFL region shown in
Fig.\ \ref{fig:2}.\cite{Kirkpatrick_Belitz_2010} Although it is not an ordered
phase, such a state has much in common with the A-phase and we will discuss it
in the context of the ordered phases.

\subsection{$O(\gso^4)$: Crystal-field effects}
\label{subsec:II.E}

To fourth order in $\gso$, we need to take into account the crystal-field terms
shown in Eq.\ (\ref{eq:1.5a}). This makes the saddle-point equations very
complicated, and no exact solution is known. We therefore take a variational
approach by inserting Eq.\ (\ref{eq:2.1a}) into the action and minimizing with
respect to the parameters of the {\it ansatz}. Of all the members of the class
of functions represented by Eqs.\ (\ref{eqs:2.1}) this will yield the one with
the lowest free energy.

By writing Eq.\ (\ref{eq:1.5a}) we have fixed the coordinate system by choosing
the crystallographic axes to be the $x,y,z \equiv 1,2,3$ axes. We thus are no
longer free to choose the direction of ${\bm H}$, ${\bf m}_0$, or ${\bm q}$.
For simplicity, we will consider only the case of a magnetic field along the
$z$-axis: ${\bm H} = (0,0,H)$. ${\hat{\bm q}} \equiv {\bm q}/q$ we parameterize
in terms of angles $\vartheta$ and $\varphi$ as follows,
\bse
\label{eqs:2.10}
\be
{\hat{\bm q}} = (\sin\vartheta\,\cos\varphi, \sin\vartheta\,\sin\varphi,
\cos\vartheta) \equiv (\beta_1, \beta_2, \beta_3),
\label{eq:2.10a}
\ee
with $\sum_{i=1}^3 \beta_i^2 = 1$. This leaves one free parameter for
${\hat{\bm e}}_+$, namely, an azimuthal angle $\varphi_e$:
\bea
{\hat{\bm e}}_+ &=& (\cos\vartheta\,\cos\varphi,\sin\varphi_e -
\sin\varphi\,\cos\varphi_e, \nonumber\\
&&\hskip 4pt \cos\vartheta\ \sin\varphi\,\sin\varphi_e +
\cos\varphi\,\cos\varphi_e,
\nonumber\\
&& -\sin\vartheta\,\sin\varphi_e).
\label{eq:2.10b}
\eea
This uniquely determines ${\hat{\bm e}}_- = {\hat{\bm q}}\times{\hat{\bm
e}}_+$. Finally, ${\bm m}_0$ in general needs to be decomposed into components
parallel and perpendicular, respectively, to ${\bm
q}$.\cite{Plumer_Walker_1981} However, while a perpendicular component can lead
to a slightly lower free energy, it has no qualitative effects on the structure
of the phase diagram, and we therefore restrict our ansatz to
\be
{\bm m}_0 = m_0\, {\hat{\bm q}}.
\label{eq:2.10c}
\ee
\ese
We further assume that the system is sufficiently close to a second order or
weakly first order phase transition that one can neglect the last term in Eq.\
(\ref{eq:1.5a}). With these approximations, the free energy does not depend on
the angle $\varphi_e$ and is completely parameterized in terms of six
parameters, namely: two amplitudes, $m_0$ and $m_1 = [(m_1^+)^2 +
(m_1^-)^2]^{1/2}/\sqrt{2}$, the polarization parameter $\pi$, the modulus $q$,
and the two direction angles $\vartheta$ and $\varphi$. We find
\bse
\label{eqs:2.11}
\begin{widetext}
\bea
f &=& \frac{1}{2}\,\delta t\,m_0^2 + \frac{r}{4}\,(m_0^2 + m_1^2)
 + \frac{u}{4}\,(m_0^2 + m_1^2)^2 +
\frac{1}{2}\,m_1^2\left[\delta t + a q^2 - cq +
\frac{1}{2}\,cq(\delta\pi)^2\right] - H m_0\,\cos\vartheta
\nonumber\\
&& + \frac{b}{4}\,m_1^2\,q^2\,B_{\text{s}}(\vartheta,\varphi)
   + \frac{b_1}{4}\,m_1^2\,q^2\,B_{1\text{s}}(\vartheta,\varphi)
   - \frac{b}{4}\,m_1^2\,q^2\,\delta\pi\,B_{\text{a}}(\vartheta,\varphi)
   - \frac{b_1}{4}\,m_1^2\,q^2\,\delta\pi\,B_{1\text{a}}(\vartheta,\varphi) +
O(\gso^8).
\label{eq:2.11a}
\eea
\end{widetext}
Here we have defined $r = t - \delta t$, with $\delta t$ arbitrary at this
point. We also have made use of the fact that we know, from Sec.\
\ref{subsec:II.B}, that the physical solution has the property $\pi = 1 +
\delta\pi$ with $\delta\pi = O(\gso^2)$, and have expanded in powers of
$\delta\pi$. The angle-dependent functions in Eq.\ (\ref{eq:2.11a}) are
\bea
B_{\text{s,a}}(\vartheta,\varphi) &=& B^+(\vartheta,\varphi) \pm
B^-(\vartheta,\varphi), \nonumber\\
B_{1\text{s,a}}(\vartheta,\varphi) &=& B_1^+(\vartheta,\varphi) \pm
B_1^-(\vartheta,\varphi),
\label{eq:2.11b}
\eea
where
\bea
B^+(\vartheta,\varphi) &=& 2 \sin^2\vartheta\, \sin^2\varphi\, \cos^2\varphi,
\nonumber\\
B^-(\vartheta,\varphi) &=& \sin^2\vartheta\, \cos^2\vartheta\,(1 +
\sin^4\varphi + \cos^4\varphi),
\nonumber\\
B_1^+(\vartheta,\varphi) &=& \sin^2 \vartheta\,\cos^4 \varphi + \cos^2
\vartheta\,\sin^2 \varphi,
\nonumber\\
B_1^-(\vartheta,\varphi) &=& \sin^4 \vartheta\,\sin^2 \varphi + \cos^4
\vartheta\,\cos^2\varphi \nonumber\\
&& + \sin^2\vartheta\,\cos^2\vartheta\, \sin^2\varphi\, \cos^2\varphi.
\label{eq:2.11c}
\eea
\ese

We now need to minimize the free energy with respect to the six parameters. We
will first consider the case $H=0$ to understand the pinning of the helix by
the crystal-field terms, and then determined the effects of a magnetic field.
Furthermore, in order to keep the discussion transparent we will initially
restrict ourselves to an {\it ansatz} with a circular polarization, $\delta\pi
= 0$. This suffices to understand the existence of the critical field $\Hcone$.
We will then generalize the {\it ansatz} to allow for a non-circular
polarization and show that this leads to a splitting of the transition at
$\Hcone$, with a first order transition from a circularly polarized helix to an
elliptically polarized one at a critical field $\Hcone' < \Hcone$ preceding the
alignment transition at $\Hcone$.

\subsubsection{Pinning of the helix}
\label{subsubsec:II.E.1}

We consider the system at $H=0$ and initially restrict our {\it ansatz} to the
case of circular polarization, $\delta\pi = 0$. The remaining angular
dependence in the free energy, Eq.\ (\ref{eq:2.11a}), is contained in the
functions $B_{\text{s}}$ and $B_{1\text{s}}$. Minimizing with respect to
$\varphi$ we find that there are two cases.
\begin{description}
\item[{\it Case (1):}] $\varphi = 0,\pi/2,\pi,3\pi/2$. This implies
${\hat{\bm q}} = (0,\pm\sin\vartheta,\cos\vartheta)$ or
$(\pm\sin\vartheta,0,\cos\vartheta)$, and $B_{\text{s}} = \frac{1}{2}\,\sin^2
2\vartheta$, $B_{1\text{s}} = \sin^2\vartheta + \cos^4\vartheta$. Minimizing
with respect to $\vartheta$ leads to two subcases:
\par%
\begin{description}
\item[{\it Case (1a):}] $2b < b_1$. The free energy is minimized by $\vartheta =
\pi/4, 3\pi/4, 5\pi/4, 7\pi/4$, which implies ${\hat{\bm q}} =
(1,0,1)/\sqrt{2}$ or equivalent.
\smallskip%
\par%
\item[{\it Case (1b):}] $2b > b_1$. The free energy is minimized by $\vartheta = 0,
\pi/2, \pi, 3\pi/2$, which implies ${\hat{\bm q}} = (1,0,0)$ or equivalent.
\end{description}
\item[{\it Case (2):}] $\varphi = \pi/4, 3\pi/4, 5\pi/4, 7\pi/4$. This implies
${\hat{\bm q}} = (\sin\vartheta/\sqrt{2},\sin\vartheta/\sqrt{2},\cos\vartheta)$
or equivalent, and $B_{\text{s}} = 1 - \sin^4\vartheta/2 - \cos^4\vartheta$,
$B_{1\text{s}} = 1 - \sin^2\vartheta + 3\sin^4\vartheta/4$. Minimizing with
respect to $\vartheta$ yields
\begin{description}
\item[{\it Case (2a):}] $2b < b_1$. The free energy is minimized by $\vartheta =
\pm\arcsin\sqrt{2/3}$, which implies ${\hat{\bm q}} = (1,1,1)/\sqrt{3}$ or
equivalent.
\item[{\it Case (2b):}] $2b > b_1$. The free energy is minimized by $\vartheta = 0,
\pi$, which implies ${\hat{\bm q}} = (0,0,\pm 1)$.
\end{description}
\end{description}

By comparing the resulting free energies for these cases we see that Case (1b)
provides the minimum for $b > b_1/2$, whereas Case (2a) provides the minimum
for $b < b_1/2$. This is a generalization of the result obtained in Ref.\
\onlinecite{Bak_Jensen_1980}, which considered a model with $b_1=0$.

In MnSi, the pinning is observed to be in the $\langle
1,1,1\rangle$-directions, which implies $b < b_1/2$, and we will focus on this
case from now on. Minimizing the free energy with respect to $q$, we find
\be
q = \frac{c/2}{a + (b + b_1)/3},
\label{eq:2.12}
\ee
which generalizes Eq.\ (\ref{eq:2.6b}). Choosing $\delta t = cq/2$ and
minimizing with respect to $m_1$ we finally have
\bse
\label{eqs:2.13}
\be
f = -r^2/4u,
\label{eq:2.13a}
\ee
where
\be
r = t - \delta t,
\label{eq:2.13b}
\ee
with
\be
\delta t = \frac{c^2/4}{a + (b + b_1)/3}.
\label{eq:2.13c}
\ee
\ese
These results are valid for $H=0$ and $b < b_1/2$.

\subsubsection{The alignment transition, and the critical field $\Hcone$}
\label{subsubsec:II.E.2}

For $H>0$ we expect the pitch vector to move away from $(1,1,1)$ towards
$(0,0,1)$. The calculation proceeds as for $H=0$, except that now the
minimization with respect to $\vartheta$ yields an $H$-dependent result. For
Case (2a) we find
\bse
\label{eqs:2.14}
\be
\beta_3 \equiv \cos\vartheta = \begin{cases} 1 & \text{if $H\geq\Hcone$}\\
                                             \frac{1}{3}\left(1 +
                                             2H^2/\Hcone^2\right) & \text{if $H
                                             < \Hcone$},
                               \end{cases}
\label{eq:2.14a}
\ee
where (remember $r<0$ and $b<b_1/2$ in the ordered phase for Case (2a))
\be
\Hcone^2 = r(b-b_1/2)(\delta t)^2/ua.
\label{eq:2.14b}
\ee
\ese
The helical pitch vector ${\hat{\bm q}}$ thus moves continuously along the
shortest path from its initial value, $(1,1,1)/\sqrt{3}$ at $H=0$, to $(0,0,1)$
at $H = \Hcone$, and remains in that position for $H > \Hcone$. There thus is a
second order transition at $H = \Hcone$ \cite{Plumer_Walker_1981} that we refer
to as the alignment transition. An inspection shows that Case (1b) has a larger
free energy for all $H < \Hcone$.

For reference in the next subsection we mention that if one expands the free
energy for small values of $\vartheta$, and looks for the instability of the
solution with $\vartheta=0$, one finds that the latter occurs at $H = \Hcone$,
as expected.

\subsubsection{The polarization transition, and the critical field $\Hcone'$}
\label{subsubsec:II.E.3}

The circular polarization {\it ansatz} we have used so far explains the two
critical fields $\Hcone$ and $\Hctwo$ observed in MnSi. However, the solution
obtained in this way misses a qualitative feature, as was first pointed out in
Ref.\ \onlinecite{Walker_1989} on symmetry grounds. Since in general $b\neq
b_1$, the crystal-field contribution to the action, Eq.\ (\ref{eq:1.5a}), is
not invariant under $x \leftrightarrow y$. As a result, there is no reason for
the $x$ and $y$-components of ${\hat{\bm q}}$ to become nonzero at the same
value of $H$ as $H$ is lowered from above, yet the solution constructed in the
previous subsection has this property. Clearly, this is a result of the fact
that our {\it ansatz} with circular polarization possesses cubic symmetry,
while the action does not. In general, one therefore expects two separate
transitions in the vicinity of $\Hcone$; one where the $x$-component of
${\hat{\bm q}}$ becomes nonzero, and a separate one where the $y$-component
becomes nonzero. We now show that this expectation is borne out if we allow for
a non-circular polarization of the helix, which breaks the cubic symmetry of
the {\it ansatz}.

Consider the full Eq.\ (\ref{eq:2.11a}), allowing for $\delta\pi \neq 0$.
Minimizing with respect to $\varphi$ we see that there are two distinct cases.
\begin{description}
\item[{\it Case (1):}] $\varphi = 0,\pi/2,\pi,3\pi/2$ and $\delta\pi$
arbitrary. Minimizing with respect to $\delta\pi$ yields
\bse
\label{eqs:2.15}
\be
\delta\pi(\vartheta) = \frac{q}{2c}\,\left[b\,B_{\text{a}}(\vartheta) +
b_1\,B_{1\text{a}}(\vartheta)\right].
\label{eq:2.15a}
\ee
That is, the polarization is in general elliptical. We have $B^+ = 0$,
$B^- = 2\sin^2\vartheta\cos^2\vartheta$, which leads to
\bea
B_{\text{s,a}} &=& \pm\frac{1}{2}\,\sin^2 2\vartheta,
\label{eq:2.15b}\\
B_{1\text{s}} &=& \sin^2\vartheta + \cos^4\vartheta.
\label{eq:2.15c}
\eea
\ese
Considering $B_{1\text{a}}$ we find two subcases. The first one is
\begin{description}
\item[{\it Case (1)(i):}] $\varphi = 0,\pi$, which implies
\bse
\label{eqs:2.16}
\be
{\hat{\bm q}} = (0,\pm\sin\vartheta,\cos\vartheta),
\label{eq:2.16a}
\ee
and
\be
B_{1\text{a}} = \sin^2\vartheta - \cos^4\vartheta.
\label{eq:2.16b}
\ee
\ese
\par\noindent
\item[{\rm The second one is}] $ $
\smallskip\par\noindent
\item[{\it Case (1)(ii):}] $\varphi = \pi/2, 3\pi/2$, which implies
\bse
\label{eqs:2.17}
\be
{\hat{\bm q}} = (\pm\sin\vartheta,0,\cos\vartheta),
\label{eq:2.17a}
\ee
and
\be
B_{1s,a} = \cos^2\vartheta \pm \sin^4\vartheta.
\label{eq:2.17b}
\ee
\ese
\end{description}
\item[{\rm The second case is}] $ $
\item[{\it Case (2):}] $\varphi = \pi/4, 3\pi/4, 5\pi/4, 7\pi/4$ and
\bse
\label{eqs:2.18}
\be
\delta\pi = 0.
\label{eq:2.18a}
\ee
That is, the polarization is circular. This implies
\be
{\hat{\bm q}} = \left(\frac{1}{\sqrt{2}}\,\sin\vartheta,
                      \frac{1}{\sqrt{2}}\,\sin\vartheta, \cos\vartheta\right),
\label{eq:2.18b}
\ee
and
\bea
B_{\text{s}} &=& 1 - \frac{1}{2}\,\sin^4\vartheta - \cos^4\vartheta,
\label{eq:2.18c}\\
B_{1,\text{s}} &=& 1 - \sin^2\vartheta + \frac{3}{4}\,\sin^4\vartheta
\label{eq:2.18d}
\eea
\ese
\end{description}

Now first consider the case $H=0$. Minimizing with respect to $\vartheta$ one
finds that, for $b < b_1/2$, Case (2) yields the lower free
energy.\cite{corrections_footnote} The physical solution is thus a circularly
polarized helix pinned in the $\langle 1,1,1\rangle$ directions, and the
relaxation of the condition we had imposed in Sec.\ \ref{subsubsec:II.E.1} does
not change anything. For $b > b_1/2$, the physical solution is an elliptically
polarized helix pinned in the $\langle 0,0,1\rangle$ directions.

Next we consider a solution with $\vartheta = 0$, which we expect to be stable
for sufficiently large $H$. It is easy to see that Case (1), which takes
advantage of the possibility of an elliptical polarization, has a free energy
that is lower than that of Case (2) by a term of $O(\gso^6)$ everywhere in the
conical phase, i.e., for $H < \Hctwo$. We next look for the instability of the
$\vartheta = 0$ solution at small $H$, which can be found by expanding the
action to second order in $\vartheta$ and looking for the zero of the
coefficient of the quadratic term. As expected, this instability occurs at a
field $\Hcone = O(\gso^3)$. To leading order in $\gso$, $\Hcone$ is given by
Eq.\ (\ref{eq:2.14b}). The values of $\Hcone$ for the two cases are different,
but the difference is only of $O(\gso^5)$ and is irrelevant for the following
argument.

\begin{figure}[t]
\vskip -0mm
\includegraphics[width=5.0cm]{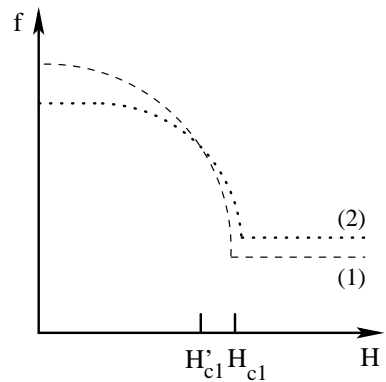}
\caption{Schematic plot of the free energy density as a function of $H$ for Case
(1) (dashed line) and Case (2) (dotted line), respectively.}
\label{fig:6}
\end{figure}

We now have the following situation. For $H < \Hcone$, the free energy of Case
(1) is lower by a term of $O(\gso^6)$. However, we know that at $H=0$ the
free energy of Case (2) is lower by a term of $O(\gso^4)$. The two solutions
cross at a field $\Hcone'$ given by
\be
\frac{\Hcone'}{\Hcone} = 1 - \frac{1}{2}\,\sqrt{\frac{3 b_1^2}{4 a \vert b -
b_1/2\vert}}
= 1 - O(\gso),
\label{eq:2.19}
\ee
see Fig.\ \ref{fig:6}.
At this value of $H$, the state will change discontinuously from an
elliptically polarized helix with a pitch vector given by either Eq.\
(\ref{eq:2.16a}) or (\ref{eq:2.17a}) to a circularly polarized one with a pitch
vector given by Eq.\ (\ref{eq:2.18b}).

We now have the following progression of phases and phase transitions as the
magnetic field is lowered from a value greater than $\Hctwo$:
\medskip\par\noindent
$H > \Hctwo$: Field-polarized state, no helix.
\smallskip\par\noindent
$H = \Hctwo \propto \gso^2$: Second-order transition to a conical state with an
elliptically polarized helix, ${\hat{\bm q}} = (0,0,1)$.
\smallskip\par\noindent
$H = \Hcone \propto \gso^3$: Second-order transition to a conical state with
elliptical polarization as above, but ${\hat{\bm q}} =
(0,\sin\vartheta,\cos\vartheta)$. $\vartheta$ increases from zero with
decreasing $H$.
\smallskip\par\noindent
$H = \Hcone' = \Hcone [1 - O(\gso)]$: First-order transition to a conical state
with circular polarization and ${\hat{\bm q}} =
(\frac{1}{\sqrt{2}}\,\sin\vartheta, \frac{1}{\sqrt{2}}\,\sin\vartheta,
\cos\vartheta)$. $\vartheta$ increases from its value at $\Hcone'$ with
decreasing $H$.
\smallskip\par\noindent
$H = 0$: System reaches helical state with circular polarization and ${\hat{\bm q}}
=(1,1,1)/\sqrt{3}$.
\medskip\par\noindent

\begin{figure}[t]
\vskip -0mm
\includegraphics[width=6.0cm]{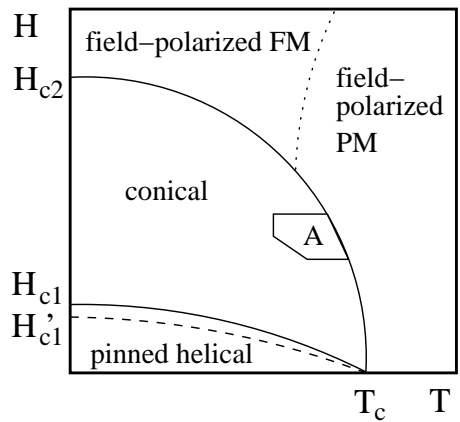}
\caption{Phase diagram of MnSi in the $H$-$T$ plane showing the predicted
 polarization transition at $\Hcone'$ (first order, dashed line) in addition to
 the second-order transitions at $\Hcone$ and $\Hctwo$. The nature of the transition
 from the conical phase to the A-phase, or the A-phase to the paramagnetic phase,
 has not been investigated so far. The dotted line denotes the crossover from a
 field-polarized paramagnet to a field-polarized ferromagnet.}
\label{fig:7}
\end{figure}

The phase diagram is thus predicted to have the structure shown in Fig.\
\ref{fig:7}, with the second order alignment transition at $\Hcone$ followed by
a first-order polarization (and re-alignment) transition at $\Hcone'$. The
latter has so far not been observed experimentally.
It is an explicit realization of the type of transition first predicted by
Walker\cite{Walker_1989} on symmetry grounds. It needs to be stressed that the
detailed features of this transition are restricted by our {\it ansatz}, Eqs.\
(\ref{eqs:2.1}); the states obtained are not true minima of the action.
However, the basic physical idea, which is expected to be realized in the true
ground state as well, is as follows. In the conical phase, where the helical
pitch vector is aligned with the magnetic field, the system can take advantage
of the lack of cubic symmetry of the action, Eq.\ (\ref{eq:1.5a}), by forming a
helix with a non-circular polarization. This leads to an energy gain, compared
to a circularly polarized state, of order $\gso^6$. However, it forces the
pitch vector into either the $y=0$ or $x=0$ plane, i.e., Case (1) above. At low
fields, these states have a free energy that is larger by a term of $O(\gso^4)$
than the states with the pitch vector along a cubic diagonal, which requires a
circular polarization. The competition between these two effects leads to the
first order transition at $\Hcone$, where both the polarization and the
orientation of the pitch vector change discontinuously.

\subsubsection{Pinning of the A-crystal and the NFL region}
\label{subsubsec:II.E.4}

Experimentally, the A-crystal is found to be pinned very weakly, and
theoretical considerations conclude that the pinning potential is only of
$O(\gso^6)$.\cite{Muhlbauer_et_al_2009} This is a consequence of the six-fold
rotation symmetry of the A-crystal. Similarly one expects the (average)
direction of the fluctuating defect lines in the NFL region to be pinned only
very weakly.

\subsection{Beyond classical mean-field theory}
\label{subsec:II.F}

We have treated the phase transitions that we have discussed in this section
within classical mean-field theory, and the question arises what changes will
result from the consideration of fluctuations, classical or quantum. One
example of the effects of fluctuations is the tricritical point and the
associated tricritical wings shown in Fig. \ref{fig:3}, which result from
quantum fluctuations that can be taken into account within a renormalized
mean-field theory.\cite{Belitz_Kirkpatrick_Vojta_1999,
Belitz_Kirkpatrick_Vojta_2005} Elsewhere in the phase diagram, fluctuations are
also of qualitative importance. Consider, for instance, the transition from the
conical phase to the field-polarized phase at the critical field $\Hctwo$, see
Fig.\ \ref{fig:1}. The transition is characterized by the vanishing of the
one-dimensional order parameter $m_1$, the amplitude of the helix, see Eqs.\
(\ref{eqs:2.6}). The action is invariant under $m_1 \to -m_1$, and we therefore
expect this transition at nonzero temperature to be in the universality class
of the classical Ising model.\cite{Ma_1976} At zero temperature the statics and
the dynamics will couple,\cite{Hertz_1976, Sachdev_1999} and one expects the
quantum phase transition to be described by Hertz's model\cite{Hertz_1976} with
a scalar order parameter.

\section{Goldstone modes}
\label{sec:III}

Goldstone's theorem states that if a continuous symmetry described by a Lie
group $G$ is spontaneously broken, with the remaining subgroup in the
broken-symmetry phase being $H$ (not to be confused with the magnetic field),
then in the ordered phase there are $n$ soft or massless modes, with $n$ equal
to the dimensionality of the coset space $G/H$.\cite{Zinn-Justin_1996} All of
the ordered states discussed in Sec.\ \ref{sec:II} break a continuous symmetry,
and therefore there must be one or more soft modes in the ordered
phase.\cite{translation_footnote} The number and functional form of the
Goldstone modes in the various phases can be determined from general arguments.
In this section we will do so, and we will augment these arguments by explicit
calculations in some cases where this is feasible. Via a coupling to the
conduction electrons, the Goldstone modes have interesting consequences for
various observables; this is the topic of Sec.\ \ref{sec:IV}.

\subsection{$O(\gso^0)$: Ferromagnons}
\label{subsec:III.A}

To zeroth order in $\gso$ the system is a ferromagnet, see Sec.\
\ref{subsec:II.A}. The relevant symmetry is the rotation group, $G = O(3)$,
which in the ordered state is spontaneously broken to $H = O(2)$. Hence there
are $\text{dim}(O(3)/O(2)) = 2$ Goldstone modes, the well-known ferromagnons
(FM). Their dispersion relation for small wave numbers is
\be
\omega_{\,\text{FM}}({\bm k}) = D\,{\bm k}^2,
\label{eq:3.1}
\ee
with $D$ the spin wave stiffness, which depends on the magnetization and
vanishes as the magnetization goes to zero.

The easiest way to derive this result is to consider a nonlinear
sigma-model-version of the Heisenberg action.\cite{Zinn-Justin_1996} Neglecting
fluctuations of the amplitude of the magnetization, which can be shown to be
massive, one parameterizes the order parameter
\be
{\bm M}({\bm x}) = m_0 \left(\begin{array}{c} \pi_1({\bm x}) \\
                                              \pi_2({\bm x}) \\
                                              \sqrt{1 - {\bm\pi}^2({\bm x})}
                             \end{array}\right)
\label{eq:3.2}
\ee
and expands the action to bilinear order in $\pi_{1,2}$. If we neglect the
small relativistic term with coupling constant $d$ in Eq.\ (\ref{eq:1.3}), the
resulting quadratic form has two identical eigenvalues
\be
\lambda = \frac{m_0}{2}\,(a\,{\bm k}^2 + H)
\label{eq:3.3}
\ee
For $H=0$, one has $\lambda({\bm k} \to 0) \to 0$, which reflects the two
Goldstone modes. Physically, the homogeneous transverse magnetic susceptibility
diverges. This is the static manifestation of the spontaneously broken
continuous symmetry. Determining the dynamics requires an additional step. One
either needs to solve an appropriate Langevin equation within a classical
context,\cite{Ma_1976} or treat the problem quantum
mechanically.\cite{Hertz_1976, Belitz_et_al_1998} Either way one obtains Eq.\
(\ref{eq:3.1}) with $D \propto m_0$.

\subsection{$O(\gso^2)$: Helimagnons}
\label{subsec:III.B}

When the spin-orbit coupling is taken into account, we have the various phases
involving helical spin textures discussed in Sec.\ \ref{subsec:II.A}. For all
of these phases the relevant symmetry is the translational one. Let $T$ be the
Lie group of one-dimensional translations. Then the action is invariant under
$G = T \otimes T \otimes T \equiv T^3$.

\subsubsection{Symmetry arguments}
\label{subsubsec:III.B.1}

Consider the aligned conical state (ACS) discussed in Sec.\ \ref{subsec:II.B},
from which the unpinned helical phase emerges as a special case at $H=0$. $T^3$
is broken down to $T^2$ (translations in the two directions perpendicular to
${\hat{\bm q}} = {\hat{\bm H}}$), so we expect one Goldstone mode. At $H=0$,
one expects the soft fluctuations in the ordered phase to be phase fluctuations
of the helix, and one might naively expect them to be governed by an action $S
= \int d{\bm x}\ ({\bm\nabla}\phi({\bm x}))^2$, with $\phi$ the phase variable,
which would lead to a soft eigenvalue proportional to ${\bm k}^2$. However, the
ordered state is also invariant under rotations of ${\bm q}$, which can be
written as a phase fluctuation with a nonvanishing gradient, so this cannot be
the correct answer. The lowest-order term allowed by rotational symmetry that
involves the gradients perpendicular to ${\bm q}$ is of the form
$({\bm\nabla}_{\perp}^2 u)^2$, with $u$ a generalized phase variable, and this
leads to an eigenvalue, or inverse order parameter susceptibility, proportional
to the term under the square root in Eq.\
(\ref{eq:1.1}).\cite{Liquid_Xtals_footnote} The dynamics again require
additional considerations, which lead to a resonance frequency that is
proportional to the square root of the inverse susceptibility, unlike the
ferromagnetic case, and this is expressed in Eq.\
(\ref{eq:1.1}).\cite{dynamics_footnote} These results were first obtained by
means of an explicit calculation for both classical and quantum helimagnets in
Ref.\ \onlinecite{Belitz_Kirkpatrick_Rosch_2006a}.

A magnetic field breaks the rotational invariance, so a ${\bm
k}_{\perp}^2$-term will be present in the dispersion relation. The prefactor is
expected to be analytic in $H$, and we thus expect for the dispersion relation
of this ``conimagnon'', the Goldstone model of the ACS,
\be
\omega_{\text{ACS}}({\bm k}) = \sqrt{c_{||}\,k_{||}^2 + c_{\perp}'\,{\bm
k}_{\perp}^2 + c_{\perp}\,{\bm k}_{\perp}^4},
\label{eq:3.4}
\ee
with $c_{\perp}' \propto H^2$.

\subsubsection{Model calculation}
\label{subsubsec:III.B.2}

For the action to $O(\gso^2)$, Eqs.\ (\ref{eqs:2.6}, \ref{eqs:2.7}) constitute
an exact saddle-point solution, and we can perform an explicit calculation of
the Goldstone mode. A complete parameterization of fluctuations about the
saddle point can be written
\bea
{\bm M}({\bm x}) &=& \left(m_0 + \delta m_0({\bm x})\right)
   \left(\begin{array}{c} \pi_1({\bm x}) \\
                          \pi_2({\bm x}) \\
                          \sqrt{1 - {\bm\pi}^2({\bm x})}
         \end{array}\right)
\nonumber\\
&& + \frac{m_1 + \delta m_1({\bm x})}{\sqrt{1 + \psi^2({\bm x})}}
   \left(\begin{array}{c} \cos(qz + \varphi_0({\bm x})) \\
                          \sin(qz + \varphi_0({\bm x})) \\
                          \psi({\bm x})
         \end{array}\right)\ .
\nonumber\\
\label{eq:3.5}
\eea
The first term is the nonlinear sigma model for the homogeneous magnetization
from Sec.\ \ref{subsec:III.A}, and the second one parameterizes fluctuations of
the helix in terms of an amplitude modulation, a phase $\varphi_0$, and a third
component $\psi$. The amplitude fluctuations one expects to be massive, and an
explicit calculation confirms this, so we drop $\delta m_0$ and $\delta m_1$.
The field $\psi({\bm x})$ is conveniently written
\be
\psi({\bm x}) = \varphi_{+}({\bm x})\,\cos({\bm q}\cdot{\bm x}) +
\varphi_{-}({\bm x})\,\sin({\bm q}\cdot{\bm x}),
\label{eq:3.6}
\ee
which ensure that $\varphi_{\pm}$ and $\varphi_0$ at zero wave number both
correspond to ${\bm M}$ at wave number $q$. Double counting is avoided by
restricting the theory to wave numbers small compared to $q$. If we use Eq.\
(\ref{eq:3.5}) in the action to $O(\gso^2)$, Eqs.\
(\ref{eq:1.3},\ref{eq:1.4a}), and expand to bilinear order in the fluctuations
$(\phi_1, \phi_2, \phi_3, \phi_4, \phi_5) \equiv (\varphi_0, \varphi_+,
\varphi_-, \pi_1, \pi_2)$, we obtain a Gaussian action
\be
{\cal A}^{(2)} = \frac{a^2\,q^4}{2uV} \sum_{\bm k} \sum_{i=1}^{5} \phi_i({\bm
k})\,\gamma_{ij}({\bm k})\,\phi_j(-{\bm k}).
\label{eq:3.7}
\ee
A sketch of the derivation, and the explicit form of the matrix $\gamma$, are
given in Appendix \ref{app:B}. Of the five eigenvalues, one goes to zero for
${\bm k} \to 0$, in agreement with the expectation from the symmetry arguments
given above. It takes the form
\bse
\label{eqs:3.8}
\be
\lambda_1 = \alpha\,{\hat k}_z^2 + \beta\,{\hat {\bm k}}_{\perp}^2 +
\delta\,{\hat {\bm k}}_{\perp}^4,
\label{eq:3.8a}
\ee
with ${\hat {\bm k}} = {\bm k}/q$, and coefficients
\bea
\alpha &=& {\hat m}_1^2\ ,
\label{eq:3.8b}\\
\beta &=& \frac{{\hat m}_0^2\, {\hat m}_1^2}{(1 + {\hat m}_1^2 + {\hat m}_0^2)}\ ,
\label{eq:3.8c}\\
\delta &=& \frac{1}{2}\,{\hat m}_1^2\, \frac{(1 + {\hat m}_1^2)^3 - {\hat
m}_0^2(1 + {\hat m}_1^4) + 2 {\hat m}_0^4 {\hat m}_1^2}{(1 + {\hat m}_1^2 +
{\hat m}_0^2)^3}\ . \nonumber\\
\label{eq:3.8d}
\eea
\ese
Here we have defined ${\hat m}_{0,1}^2 = u\,m_{0,1}^2/a q^2$. This result is
consistent with Eq.\ (\ref{eq:3.4}). For $H=0$, which implies ${\hat m}_0 = 0$,
it reduces to the helimagnon result of Ref.\
\onlinecite{Belitz_Kirkpatrick_Rosch_2006a}. In addition, there are four
massive eigenvalues that appear in pairs. At zero wave number, they are
\bse
\label{eqs:3.9}
\bea
\lambda_2 = \lambda_3 \equiv \lambda_{\varphi} = {\hat m}_1^2 [1 + {\hat m}_0^2
+ O({\hat m}_0^4)],
\label{eq:3.9a}\\
\lambda_3 = \lambda_4 \equiv \lambda_{\pi} = {\hat m}_0^2 [1 + {\hat m}_1^2] +
O({\hat m}_0^4).
\label{eq:3.9b}
\eea
\ese
We recognize $\lambda_{\varphi}$ as representing the massive helimagnon
modes,\cite{Belitz_Kirkpatrick_Rosch_2006a} modified by the the presence of
$m_0$, and $\lambda_{\pi}$ as the massive (due to the presence of a magnetic
field) ferromagnons, Eq.\ (\ref{eq:3.3}), modified by the presence of $m_1$.

\subsection{$O(\gso^2)$: Skyrmionic Goldstone modes in the A-phase and the NFL
 region}
\label{subsec:III.C}

The helical states that have been proposed as candidates for the A-phase and
were discussed in Sec.\ \ref{subsec:II.E} are not saddle-point solutions of the
action, which precludes a model calculation of the Goldstone modes resulting
from this type of order. However, assuming that the order is stabilized by some
mechanism, the functional form of the soft modes can still be determined by
symmetry arguments analogous to those put forward in the previous subsection.

Consider a skyrmion lattice state. The state described by Eq.\ (\ref{eq:2.9})
is invariant only under translations in one direction, viz., the direction
perpendicular to the plane of helices. The same is true for any state that is
characterized by columnar order, so this property does not depend on the
precise nature of the skyrmions. Any such state will thus have
$\text{dim}(T^3/T) = 2$ Goldstone modes. This was to be expected: Since the
skyrmions form a two-dimensional lattice, there should be two generalized
phonon modes, namely, a compression mode and a shear mode. In zero magnetic
field, the energy would still be invariant under global rotations of the
skyrmion lattice. Hence, the soft eigenvalue can have no $k_z^2$ contribution.
For $H\neq 0$ this is no longer true, and we thus expect
\be
\lambda = \alpha\,{\bm k}_{\perp}^2 + \beta\,k_z^4 + \gamma\,k_z^2,
\label{eq:3.10}
\ee
for the soft eigenvalue, with $\gamma \propto H^2$, and
\be
\omega_{\,\text{SL}}({\bm k}) = \sqrt{c_{\perp}\,{\bm k}_{\perp}^2 +
c_z'\,k_z^2 + c_z\,k_z^4},
\label{eq:3.11}
\ee
for the dispersion relation, with $c_{\perp}, c_z = O(1)$ and $c_z'= O(H^2)$.

If the NFL region can be interpreted as a molten A-crystal, see Sec.\
\ref{subsec:II.D} then one of the two Goldstone modes, the compression mode,
will persist as long as their is columnar structure. This is analogous to the
fact that longitudinal phonons, or ordinary sound, exist in a liquid, whereas
in a crystal one has transverse phonons or shear modes in addition. The NFL
region extends to $H=0$, where the dispersion relation of the compressional
Goldstone mode is given by Eq.\ (\ref{eq:3.11}) with $c_z'=0$ to first order in
$\gso$.

\subsection{$O(\gso^4)$: Effects of the crystal-field terms}
\label{subsec:III.D}

\subsubsection{Symmetry arguments}
\label{subsubsec:III.D.1}

Now consider the crystal-field terms in the action that first appear at
$O(\gso^4)$, Eq.\ (\ref{eq:1.5a}). For simplicity, let us consider the term
with coupling constant $b$. It breaks rotational invariance, which invalidates
the argument that leads to the absence of a ${\bm k}_{\perp}^2$-term in the
soft mode resonance frequency. The system must remain stable regardless of the
sign of $b$, and we thus expect for the dispersion relation of the helimagnons
in the pinned helical state
\be
\omega_{\,\text{HM}}({\bm k}) = \sqrt{c_{||} k_{||}^2 + c_{\perp}''{\bm
k}_{\perp}^2 + c_{\perp} {\bm k}_{\perp}^4},
\label{eq:3.12}
\ee
with $c_{\perp}'' \propto \vert b\vert$, which replaces Eq.\ (\ref{eq:1.1}).
For the more general model given by Eqs.\ (\ref{eq:1.2}) - (\ref{eqs:1.5}),
$b_1$, $v$, and $H$ will also contribute to the elastic constant $c_{\perp}''$.

\subsubsection{Model calculation}
\label{subsubsec:III.D.2}

We now check this expectation by means of an explicit calculation. For the
model with only the first of the crystal-field terms present, we have an exact
saddle-point solution, namely,
\bse
\label{eqs:3.13}
\be
{\bm M}_{\text{sp}}({\bm x}) = m_1\,\left[{\hat{\bm e}}_+\,\cos({\bm
q}\cdot{\bm x}) + {\hat{\bm e}}_-\,\sin({\bm q}\cdot{\bm x})\right],
\label{eq:3.13a}
\ee
with ${\hat{\bm e}}_+$, ${\hat{\bm e}}_-$, and ${\bm q}$ a {\it dreibein},
\be
m_1 = \sqrt{-r/u},
\label{eq:3.13b}
\ee
where
\be
r = t - cq/2,
\label{eq:3.13c}
\ee
and
\bea
{\bm q} &=& q \begin{cases} (1,1,1)/\sqrt{3} & \text{if $b<0$}
                            \\
                            (1,0,0)          & \text{if $b>0$},
            \end{cases}
\label{eq:3.13d}\\
q &=& \begin{cases} c/2(a+b/3) & \text{if $b<0$}
                    \\
                    c/2a       & \text{if $b>0$}.
      \end{cases}
\label{eq:3.13e}
\eea
\ese
The parameterization of fluctuations about this state is given by the second
term in Eq.\ (\ref{eq:3.5}:
\bea
{\bm M}({\bm x}) &=& \left(m_1 + \delta m_1\right) \bigl[{\hat{\bm
e}}_+\,\cos({\bm q}\cdot{\bm x} + \varphi_0({\bm x}))
\nonumber\\
&& + {\hat{\bm e}}_-\,\sin({\bm q}\cdot{\bm x} + \varphi_0({\bm x})) +
{\hat{\bm q}}\, \psi({\bm x})\bigr],
\label{eq:3.14}
\eea
with $\psi({\bm x})$ given by Eq.\ (\ref{eq:3.6}). We again drop the massive
amplitude fluctuations and expand the action to quadratic order in the phase
fluctuations. The Gaussian action is of the form
\be
{\cal A}^{(2)}[\varphi_i] = \frac{m_1^2}{2} \sum_{\bm k} \sum_{\alpha=0,\pm}
\varphi_{\alpha}({\bm k})\, \Gamma_{\alpha\beta}({\bm k})\,
\varphi_{\beta}({\bm k}).
\label{eq:3.15}
\ee
The explicit form of the matrix $\Gamma$ depends on the sign of $b$; it is
given explicitly in Appendix \ref{app:C}. An inspection of the eigenvalues
shows that in either case there is one eigenvalue $\lambda$ that vanishes as
${\bm k} \to 0$ and hence represents the Goldstone mode, as expected from the
symmetry argument given above. To order $k_{||}^2$ and $k_{\perp}^4$ we find
\be
\lambda = \begin{cases} a\,k_{||}^2 + \frac{1}{2}\,b\,{\bm k}_{\perp}^2 +
                        + \frac{1}{2}\,(a+b)\,{\bm k}_{\perp}^4/q^2 & \hskip -50pt
                        \text{if $b>0$},
                        \\
                        (a + b/3)\,k_{||}^2 + \frac{2}{3}\,\vert b\vert\,
                        {\bm k}_{\perp}^2 + \frac{1}{2}\,(a+b+O(\gso^4))\,
                        {\bm k}_{\perp}^4/q^2
                        \\
                        & \hskip -50pt \text{if $b<0$}.
          \end{cases}
\label{eq:3.16}
\ee
In the case $b<0$ we have neglected terms proportional to $b^2 = O(\gso^4)$ in
the prefactor of ${\bm k}_{\perp}^4/q^2$. This result agrees with the
functional form obtained by symmetry arguments alone, Eq.\ (\ref{eq:3.1}), and
for $b=0$ it correctly reduces to the result for the isotropic model, Ref.\
\onlinecite{Belitz_Kirkpatrick_Rosch_2006a} and Eq.\ (\ref{eq:1.1}).

\subsubsection{Generalized helimagnons}
\label{subsubsec:III.D.3}

We can now summarize the results for the single-helix phases discussed above as
follows. In the helical and conical phases, including pinning effects, there is
a single Goldstone mode with a resonance frequency
\be
\omega_0({\bm k}) = \sqrt{c_{||}\,k_{||}^2 + {\tilde c}_{\perp}\,{\bm
k}_{\perp}^2 + c_{\perp}\,{\bm k}_{\perp}^4}\ ,
\label{eq:3.17}
\ee
with ${\tilde c}_{\perp} = O(H^2,\gso^2)$ small compared to $c_{||}$ and
$c_{\perp}$. This comprises Eqs.\ (\ref{eq:3.4}) and (\ref{eq:3.12}).

\subsubsection{A-phase, and NFL region}
\label{subsubsec:III.D.4}

In the A-phase, pinning effects are weaker than in the helical phase due to the
hexagonal nature of the skyrmion lattice, see Sec.\ \ref{subsubsec:II.E.4}
above. When this weak pinning is taken in to account, the Goldstone mode is
thus given by Eq.\ (\ref{eq:3.11}) with $c_z' = O(\gso^6, H^2)$. While in MnSi
the A-phase is observed only in an external magnetic field, there is no
intrinsic reason why in some other system it could not be stable in a zero
field. The Goldstone modes in such a system would be given by Eq.\
(\ref{eq:3.11}) with an extremely small $c_z'$. By the same argument, in the
NFL region at $H=0$ we expect a Goldstone mode given by Eq.\ (\ref{eq:3.11})
with $c_z' = O(\gso^6)$.

\subsection{Summary of Goldstone modes, and temperature regimes}
\label{subsec:III.E}

To summarize, we have found that the (single) Goldstone mode in the pinned
helical and conical phases is given by Eq.\ (\ref{eq:3.17}). In the pinned
helical phase the parallel direction is determined by the crystal-field effects
that pin the helix; in the conical phase, it is the direction of the magnetic
field (which we have chosen to be the $z$-direction for all of our
considerations). The elastic constant ${\tilde c}_{\perp}$ is small compared to
the other elastic constants. In the pinned helical phase it is of $O(\gso^2)$,
and in the conical phase it is of $O(H^2)$. By contrast, $c_{||}$ and
$c_{\perp}$ are of $O(\gso)$ and $O(\gso^0)$, and of $O(H^0)$, respectively. In
the A-phase, there are two Goldstone modes whose dispersion relation is given
by Eq.\ (\ref{eq:3.11}). The elastic constant $c_z'$ is small of $O(H^2)$ and
$O(\gso^6)$ compared to the others. In the NFL region, the single Goldstone
mode is also given by Eq.\ (\ref{eq:3.11}), with $c_z' = O(\gso^6)$. Finally,
in the perpendicular conical state discussed in Appendix \ref{app:A} there is a
single Goldstone mode that in the absence of crystal-field effects is given by
Eq.\ (\ref{eq:A.3}). If pinning by crystal-field terms is taken into account
there is an additional term $c_y\,k_y^2$ under the square root, with $c_y =
O(\gso^2)$. All of these results are also summarized in Table \ref{tab:1} in
Sec.\ \ref{sec:V}.

The structures of the various Goldstone modes, and the fact that the various
elastic coefficients have very different magnitudes, leads to the formation of
different temperature regimes that are dominated by different physics. We now
explain this using the generalized helimagnons, Eq.\ (\ref{eq:3.17}), as an
example; the argument for the other cases works analogously. As far as the
coupling of the magnetic Goldstone mode to the electronic degrees of freedom is
concerned, the resonance frequency $\omega_0$ scales as the temperature,
$\omega_0 \sim T$. If we scale $k_{||}$ with $T/\sqrt{c_{||}}$ and $k_{\perp}$
with $\sqrt{T}/c_{\perp}^{1/4}$, we obtain
\be
\omega_0({\bm k}) = T \sqrt{k_{||}^2 + {\bm k}_{\perp}^2 {\tilde
c}_{\perp}/T\sqrt{c_{\perp}} + {\bm k}_{\perp}^4},
\label{eq:3.18}
\ee
where $k$ now denotes the scaled, dimensionless, wave number. For $T\gg {\tilde
c}_{\perp}/\sqrt{c_{\perp}}$ the symmetry-breaking ${\bm k}_{\perp}^2$ term is
negligible, and the Goldstone mode is effectively what it would be in a
rotationally invariant system. In this regime the physics is dominated by
universal hydrodynamic effects that are independent of the microscopic details
of the solid and analogous to the corresponding effects in liquid crystals. In
the opposite limit, $T\ll {\tilde c}_{\perp}/\sqrt{c_{\perp}}$, the
crystal-field effects due to the underlying ionic lattice, or the external
magnetic field, if present, dominate and the Goldstone mode has the same
functional form as (anisotropic) acoustic phonons. Due to the smallness of
${\tilde c}_{\perp}$ the universal hydrodynamic regime is sizable, and it is in
this region that the most interesting effects of the magnetic order manifest
themselves in the electronic properties of the system. This is true especially
if the Goldstone mode appears in zero magnetic field and the pinning is very
small, such as in the NFL region or a (so far hypothetical) A-phase in zero
field.

\section{Effects of Goldstone modes on electronic properties}
\label{sec:IV}

Via a coupling to the conduction electrons, the Goldstone modes derived in the
preceding section influence the electronic properties of the helical magnet. In
this section we derive the consequences for the specific heat, the
single-particle relaxation rate, and the thermal and electrical resistivities.
In all cases we consider the contribution of the Goldstone mode in isolation;
it comes in addition to all other contributions to these observables.

\subsection{Specific heat}
\label{subsec:IV.A}

Any well-defined (i.e., not overdamped) excitation with a dispersion relation
$\omega({\bm k})$ yields a contribution to the specific heat $C$ given by
\be
C(T) = \frac{\partial}{\partial T}\,\frac{1}{V} \sum_{\bm k} \omega({\bm k})\,
n_{\text{B}}(\omega({\bm k})).
\label{eq:4.1}
\ee
Here $n_{\text{B}}(x) = 1/(\exp(x/T)-1)$ is the Bose distribution function, $V$
is the system volume, and we use units such that $\hbar = \kB = 1$. This allows
one to determine the contributions to the specific heat by the various
Goldstone modes.

\subsubsection{Generalized helimagnons}
\label{subsubsec:IV.A.1}

We first consider the helical and aligned conical phase. The dispersion
relation is given by Eq.\ (\ref{eq:3.17}). Performing the integral in Eq.\
(\ref{eq:4.1}) yields
\be
C(T) = {\text{const.}}\times \begin{cases}
           T^3/\sqrt{c_z}\,{\tilde c}_{\perp} &
                 \text{if $T \ll {\tilde c}_{\perp}/\sqrt{c_{\perp}}$} \\
           T^2/\sqrt{c_z\,c_{\perp}} &
                 \text{if $T \gg {\tilde c}_{\perp}/\sqrt{c_{\perp}}$}
                             \end{cases}\quad,
\label{eq:4.2}
\ee
The universal hydrodynamic result, $C(T) \propto T^2$, was first derived in
Ref.\ \onlinecite{Belitz_Kirkpatrick_Rosch_2006b}. It is subleading to, but
distinct from, the Fermi-liquid result $C(T) \propto T + O(T^3\ln T)$. At
asymptotically low temperatures it crosses over to a $T^3$ behavior consistent
with the acoustic-phonon-like dispersion relation in either the pinned helical
phase or the aligned conical phase at asymptotically small wave numbers.

\subsubsection{A-phase}
\label{subsubsec:IV.A.2}

In the skyrmion-lattice state we find from Eq.\ (\ref{eq:3.11}) in conjunction
with Eq.\ (\ref{eq:4.1})
\be
C(T) = {\text{const.}}\times \begin{cases}
           T^{3}/c_{\perp}\,c_z' &
                 \text{if $T \ll c_z'/\sqrt{c_z}$} \\
           T^{5/2}/c_{\perp}\,c_z^{1/4} &
                 \text{if $T \gg c_z'/\sqrt{c_z}$}\ .
                             \end{cases}
\label{eq:4.3}
\ee

\subsection{Relaxation times and transport coefficients}
\label{subsec:IV.B}

The temperature dependence of the single-particle relaxation rate, as well as
that of the electrical transport relaxation rate or the electrical resistivity,
can be obtained by using the results of Refs.\
\onlinecite{Kirkpatrick_Belitz_Saha_2008a, Kirkpatrick_Belitz_Saha_2008b}. The
former paper obtained the following expression for the single-particle
relaxation rate,
\bea
\frac{1}{\tau({\bm k})} &=& 2 \int_{-\infty}^{\infty} \frac{du}{\sinh(u/T)}\
\frac{1}{V}\sum_{\bm p} V''({\bm p}-{\bm k};{\bm k},{\bm p}\,;u)
\nonumber\\
&&\hskip 80pt \times\delta(u - \omega_1({\bm p})).
\label{eq:4.4}
\eea
Here the quasiparticle momentum ${\bm k}$ is taken to be on the 1-Fermi
surface, $\omega_1({\bm k}) = 0$, with
\bse
\label{eqs:4.5}
\be
\omega_{1,2}({\bm k}) = \frac{1}{2}\,\left[\xi_{{\bm k}+{\bm q}} + \xi_{\bm k}
\pm \sqrt{(\xi_{{\bm k}+{\bm q}} - \xi_{\bm k})^2 + 4\lambda^2}\right],
\label{eq:4.5a}
\ee
which is separated from the 2-Fermi surface, $\omega_2({\bm k}) = 0$, by twice
the Stoner gap $\lambda$. Here
\be
\xi_{\bm k} = \epsilon_{\bm k} - \epsilonF,
\label{eq:4.5b}
\ee
with
\be
\epsilon_{\bm k} = {\bm k}^2/2\me + \frac{\nu}{2\me\kF^2}\,(k_x^2 k_y^2 + k_y^2
k_z^2 + k_z^2 k_x^2)
\label{eq:4.5c}
\ee
\ese
an electronic energy-momentum relation consistent with a cubic crystal. The
dimensionless parameter $\nu$ is a measure of deviations from a nearly-free
electron model. $V''$ is the spectrum of an effective potential given by
\bse
\label{eqs:4.6}
\be
V(k;{\bm p}_1,{\bm p}_2) = \frac{1}{2}\,\lambda^2\,\chi(k)\,{\tilde\gamma}({\bm
k},{\bm p}_1)\,{\tilde\gamma}(-{\bm k},{\bm p}_2),
\label{eq:4.6a}
\ee
with vertices
\bea
{\tilde\gamma}({\bm k},{\bm p}) &=& \frac{q}{8\me\lambda}\,\left[k_z +
\frac{\nu}{\kF^2}\,\left(k_z {\bm p}_{\perp}^2 + 2({\bm k}_{\perp}\cdot{\bm
p}_{\perp})p_z\right)\right]
\nonumber\\
&& \hskip 100pt + O(k^2),
\label{eq:4.6b}
\eea
and $\chi$ the soft-mode susceptibility,
\be
\chi(k) = \frac{1}{2\NF}\,\frac{q^2}{3\kF^2}\,\frac{1}{\omega_0^2({\bm k}) -
(i\Omega)^2}\ .
\label{eq:4.6c}
\ee
\ese
Here $\omega_0({\bm k})$ is the resonance frequency, which is equal to the
frequency given in Eq.\ (\ref{eq:3.4}), (\ref{eq:3.11}), (\ref{eq:3.12}), or
(\ref{eq:A.3}), depending on the phase under consideration. The electrical
transport relaxation time, which determines the electrical resistivity, is
effectively given by averaging a similar expression over the Fermi
surface,\cite{Kirkpatrick_Belitz_Saha_2008b}
\bea
\frac{1}{\tau_{\text{el}}^{\text tr}} &=& \frac{1}{\NF} \int_{-\infty}^{\infty}
\frac{du}{\sinh(u/T)}\ \frac{1}{V^2}\sum_{{\bm p},{\bm k}} \frac{({\bm p}-{\bm
k})^2}{\kF^2}
\nonumber\\
&&\hskip 0pt \times V''({\bm p}-{\bm k};{\bm k},{\bm p}\,;u)\delta(u -
\omega_1({\bm p}))\,\delta(\omega_1({\bm k})).
\nonumber\\
\label{eq:4.7}
\eea
Note the additional, compared to Eq.\ (\ref{eq:4.4}), factor of $({\bm p}-{\bm
k})^2$ under the integral in Eq.\ (\ref{eq:4.7}). This is characteristic of the
description of electrical transport in a Boltzmann approximation and leads to a
temperature dependence of the electrical resistivity that is different from
that of the single-particle relaxation rate.\cite{Wilson_1954, Ziman_1972} In
contrast, in a Boltzmann description of thermal transport this additional
factor is absent, and the temperature dependence of the thermal conductivity is
given by that of the single-particle relaxation rate.\cite{Ziman_1972}

We note that these expressions for $1/\tau({\bm k})$ and
$1/\tau_{\text{el}}^{\text{tr}}$ vanish as $q\to 0$. While they give the
leading asymptotic temperature dependence for the relaxation rates in
helimagnets, they therefore cannot be used to obtain the corresponding results
in the ferromagnetic limit. The asymptotic low-$T$ behavior of the rates in
ferromagnets is qualitatively different and briefly treated in Appendix
\ref{app:D}. In the context of helimagnets, for a non-spherical Fermi surface
(i.e., $\nu \neq 0$ in Eq.\ (\ref{eq:4.5c})), and for generic wave vectors
${\bm k}$, the temperature scaling behavior of the single-particle relaxation
rate $1/\tau \equiv 1/\tau({\bm k})$, the thermal resistivity
$\rho_{\text{th}}$, and the electrical resistivity $\rho_{\text{el}}$ can be
represented schematically by the expressions
\bse
\label{eqs:4.8}
\bea
\frac{1}{\tau} &\sim& \rho_{\text{th}} \sim \int dp_{\,||}\int d{\bm
p}_{\perp}^2\ \frac{{\bm
p}_{\perp}^2 + p_{\,||}^2}{\sinh(\omega_0({\bm p})/T)}\ \nonumber\\
&&\hskip 40pt \times\frac{\delta(\omega_0({\bm p}) - p_{\perp} -
p_{\,||})}{\omega_0({\bm p})}\ .
\label{eq:4.8a}\\
\rho_{\text{el}} &\sim& \int dp_{\,||}\int d{\bm p}_{\perp}^2\ \frac{({\bm
p}_{\perp}^2 + p_{\,||}^2)^2}{\sinh(\omega_0({\bm p})/T)}\
\nonumber\\
&&\hskip 40pt \times\frac{ \delta(\omega_0({\bm p}) - p_{\perp} -
p_{\,||})}{\omega_0({\bm p})}\ .
\label{eq:4.8b}
\eea
\ese
The resonance frequency always scales as the temperature, $\omega_0 \sim T$,
and the temperature dependence of the relaxation rates thus is determined by
how the momentum components scale with temperature.

\subsubsection{Generalized helimagnons}
\label{subsubsec:IV.B.1}

For the helical and conical phases we have, from Eq.\ (\ref{eq:3.17}), $p_{||}
\sim T$, and $p_{\perp} \sim T$ and $\sim T^{1/2}$ at asymptotically low and
intermediate temperatures, respectively. This yields the following temperature
dependence for the single-particle relaxation rate and the thermal resistivity:
\be
\frac{1}{\tau} \propto \rho_{\text{th}} \propto \begin{cases} T^3 & \text{if $T
\ll {\tilde
c}_{\perp}/\sqrt{c_{\perp}}$}\\
T^{3/2} & \text{if $T \gg {\tilde c}_{\perp}/\sqrt{c_{\perp}}$}
                                 \end{cases}\ .
\label{eq:4.9}
\ee
The corresponding result for the electrical resistivity is
\be
\rho_{\text{el}} \propto \begin{cases} T^5 & \text{if $T \ll {\tilde
c}_{\perp}/\sqrt{c_{\perp}}$}\\
T^{5/2} & \text{if $T \gg {\tilde c}_{\perp}/\sqrt{c_{\perp}}$}
                                 \end{cases}\ .
\label{eq:4.10}
\ee
In a vanishing external field, and in a temperature regime where pinning
effects are not relevant, we recover the $T^{3/2}$ and $T^{5/2}$ behavior for
$1/\tau$ and $\rho_{\text{el}}$, respectively, of Ref.\
\onlinecite{Belitz_Kirkpatrick_Rosch_2006b} and
\onlinecite{Kirkpatrick_Belitz_Saha_2008b}.

\subsubsection{A-phase}
\label{subsubsec:IV.B.2}

For a skyrmion lattice, we obtain by using Eq.\ (\ref{eq:3.11})
\be
\frac{1}{\tau} \propto \rho_{\text{th}} \propto \begin{cases} T^3 & \text{if
                                                     $T \ll c_z'/\sqrt{c_z}$}\\
T^2 & \text{if $T \gg c_z'/\sqrt{c_z}$}
                                 \end{cases}\ .
\label{eq:4.11}
\ee
for the single-particle rate, and
\be
\rho_{\text{el}} \propto \begin{cases} T^5 & \text{if $T \ll c_z'/\sqrt{c_z}$}\\
T^3 & \text{if $T \gg c_z'/\sqrt{c_z}$}
                                 \end{cases}\ .
\label{eq:4.12}
\ee
for the electrical resistivity.

\subsection{Systems with quenched disorder}
\label{subsec:IV.C}

The preceding results hold for clean systems. In the presence of quenched
disorder, elastic scattering of the conduction electrons leads to profound
effects that manifest themselves in the transport properties. One needs to
distinguish between the strong-disorder regime, where the transport is
diffusive, and the weak-disorder regime, where it is ballistic. In a Fermi
liquid, these two regimes are characterized by $T\tau \ll 1$ and $T\tau \gg 1$,
respectively.\cite{Zala_Narozhny_Aleiner_2001} In a helical magnet, the
weak-disorder or ballistic regime is characterized
by\cite{Kirkpatrick_Belitz_Saha_2008a, Kirkpatrick_Belitz_Saha_2008b}
\be
\sqrt{(\epsilonF\tau_{\text{el}})^2 T/\lambda} \gg 1,
\label{eq:4.13}
\ee
where $\tau_{\text{el}}$ is the elastic relaxation time. In this regime there
is an additional contribution $\delta (1/\tau)$ to the relaxation rates that is
qualitatively the same for both the single-particle rate and the electron
transport rate,\cite{rates_footnote} and thus provides the temperature
dependence of the corrections to both the electrical and thermal resistivities.
It was shown in Refs.\ \onlinecite{Kirkpatrick_Belitz_Saha_2008a,
Kirkpatrick_Belitz_Saha_2008b} that, for temperature scaling purposes, this
contribution can be represented by
\bea
\delta(1/\tau) &\propto& \delta\rho_{\text{el}} \propto \delta\rho_{\text{th}}
\nonumber\\
&& \hskip -40pt \sim \int du\, n_{\text{F}}(u/T) \int dp_{\,||} \int
dp_{\perp}^2\ \frac{1}{\omega_0({\bm p})}\ \delta(u - \omega_0({\bm p})).
\nonumber\\
\label{eq:4.14}
\eea

\subsubsection{Generalized helimagnons}
\label{subsubsec:IV.C.1}

From Eq.\ (\ref{eq:4.14}) we see that at temperatures where pinning effects are
not important, generalized helimagnons lead to $\delta(1/\tau) \propto T$, a
result first obtained in Ref.\ \onlinecite{Kirkpatrick_Belitz_Saha_2008b}. At
asymptotically low temperatures, characterized by $T\ll {\tilde
c}_{\perp}/\sqrt{c_{\perp}}$, one finds a $T^2$-behavior. In the pinned helical
phase the crossover temperature between these two regimes is determined by the
strength of the crystal-field effects; in the conical phase the magnetic field
also cuts off the universal hydrodynamics $T$-behavior.

\subsubsection{A-phase}
\label{subsubsec:IV.C.2}

For the A-phase, Eq.\ (\ref{eq:3.11}) yields
\be
\delta(1/\tau) \propto \begin{cases} T^2 & \text{if $T \ll c_z'/\sqrt{c_z}$}\\
T^{3/2} & \text{if $T \gg c_z'/\sqrt{c_z}$}
                                 \end{cases}\ .
\label{eq:4.15}
\ee
The pinning effects in the A-phase are very weak due to the hexagonal nature of
the skyrmion lattice, with $c_z'$ only of $O(\gso^6)$, see Secs.\
\ref{subsubsec:II.E.4} and \ref{subsubsec:III.D.4}. The size of the asymptotic
region is therefore likely to be dominated by the $H$-dependence of $c_z'$.
Whether or not the universal hydrodynamic $T^{3/2}$-behavior is observable in
the A-phase (there currently are no experimental indications that it is) would
require a detailed quantitative analysis that goes beyond the scope of the
current paper.

\subsubsection{NFL region}
\label{subsubsec:IV.C.3}

The preceding result is also of interest with respect to the non-Fermi-liquid
region shown in Fig.\ \ref{fig:2}, which is {\em not} a phase with long-range
order, but where the electrical conductivity shows a pronounced
$T^{3/2}$-behavior.\cite{Pfleiderer_Julian_Lonzarich_2001} An explanation that
has recently been proposed\cite{Kirkpatrick_Belitz_2010} is as follows. The
$T^{3/2}$-behavior derived above is a consequence of the structure of the
Goldstone modes due to columnar fluctuations, Eq.\ (\ref{eq:3.11}), in
conjunction with weak quenched disorder. In the A-phase, which displays
long-range order in the form of a skyrmion lattice, there are two such
Goldstone modes, see Sec.\ \ref{subsec:III.C}. If the NFL region can be
interpreted as a melted skyrmion lattice, then the resulting skyrmion fluid
will still have one Goldstone mode with the same structure, namely, the
compression mode mentioned in Sec.\ \ref{subsec:III.C}. Weak quenched disorder
will then still produce a contribution to the electrical resistivity, as well
as to the single-particle relaxation rate, that is proportional to $T^{3/2}$ in
a pre-asymptotic region. The NFL region is observed to extend to a vanishing
external magnetic field, so the low-$T$ boundary of the universal hydrodynamic
region is determined by the pinning effects, which are very weak, see Secs.\
\ref{subsubsec:II.E.4} and \ref{subsubsec:III.D.4}. The universal hydrodynamic
$T^{3/2}$ behavior is therefore expected to extend to very low temperatures. A
remaining question is the size of the prefactor, which in a bare theory is
expected to be small due to the long length scale set by the skyrmion lattice.
The resolution proposed in Ref.\ \onlinecite{Kirkpatrick_Belitz_2010} is that
mode-mode coupling effects drastically enhance the magnitude of the effect, in
analogy to what is believed to happen in the blue phases of liquid
crystals.\cite{Englert_et_al_2000, Longa_Ciesla_Trebin_2003, Ciesla_Longa_2004}

\section{Summary, and Conclusion}
\label{sec:V}

\begin{table*}[t]
\begin{ruledtabular}
\begin{tabular}{c c|c|c||c|c|c|c}
& & & & pinned helical & A-phase & PCS\footnotemark[1] & NFL\footnotemark[2]
\\
& & & & /conical & $({\bm H}\vert\vert{\hat{\bm
    z}})$ & $({\bm H}\vert\vert{\hat{\bm z}}\ ,\ {\bm q}\vert\vert{\hat{\bm x}})$ &
 \\
\hline\hline
& & & & & & &  \\
& & Goldstone & $\omega_0({\bm k})$ & $\sqrt{c_{||}k_{||}^2
 + {\tilde c}_{\perp} {\bm k}_{\perp}^2 + c_{\perp}{\bm k}_{\perp}^4}$ &
   $\sqrt{c_{z}'k_{z}^2 + c_{z} k_{z}^4 + c_{\perp} {\bm k}_{\perp}^2}$ &
   $\sqrt{c_x k_x^2 + c_z k_z^2 + c_{\perp} {\bm k}_{\perp}^4}$  &
   $\sqrt{c_{||} k_{||}^4 + c_{\perp}{\bm k}_{\perp}^2}$ \\
& & & & & & & \hskip -374pt \hrulefill\\
& & modes     & \# & 1 & 2 & 1 & 1 \\
& & & & & & &  \\
\hline\hline & & & & & & &
\\
& & & ${{\rm universal\ hydro-} \atop {\rm dynamic\ regime}}$ & $T^2$ &
$T^{5/2}$ & $T^2$ &
\\
& & $C(T)$ & & & & \hskip -276pt \hrulefill & $T^{5/2}$
\\
& & & ${{\rm crystal-field} \atop {\rm regime}\ (T\to 0)}$ & $T^3$ & $T^3$ &
$T^{5/2}$ &
\\
& & & & & & & \hskip -422pt \hrulefill
\\
& & $1/\tau(T)$, & ${{\rm universal\ hydro-} \atop {\rm dynamic\ regime}}$ &
$T^{3/2}$ & $T^2$
                                      & $T^{3/2}$ &
\\
\begin{sideways} \hskip -10pt clean \end{sideways} & & & & &
                                                   & \hskip -276pt \hrulefill & $T^2$
\\
& & $\rho_{\text{th}}(T)$ &${{\rm crystal-field} \atop {\rm regime}\ (T\to 0)}$
                          & $T^3$ & $T^3$ & $T^2$ &
\\
&& & & & & & \hskip -422pt \hrulefill
\\
& & & ${{\rm universal\ hydro-} \atop {\rm dynamic\ regime}}$ & $T^{5/2}$ &
$T^3$ & $T^{5/2}$ & \\
& & $\rho_{\text{el}}(T)$ & & & & \hskip -276pt \hrulefill & $T^3$\\
& & & ${{\rm crystal-field} \atop {\rm regime}\ (T\to 0)}$ & $T^5$ & $T^5$
    & $T^3$ &
\\
& & & & & &
\\
\hline\hline%
& & & & & & &
\\
& & $\delta\rho_{\text{el}}(T)$, & ${{\rm universal\ hydro-} \atop {\rm
dynamic\ regime}}$ & {$T$} & $T^{3/2}$
  & $T$ &
\\
\begin{sideways} \hskip -20pt weak \end{sideways}
  & \begin{sideways} \hskip -20pt disorder \end{sideways}
  & & & & & \hskip -276pt \hrulefill & $T^{3/2}$
\\
& & \raisebox{10pt}{$\delta\rho_{\text{th}}(T)$}
  & \raisebox{10pt}{${{\rm crystal-field} \atop {\rm regime}\footnotemark[3]}$ }
  & \raisebox{10pt}{$T^2$}
  & \raisebox{10pt}{$T^2$}
  & \raisebox{10pt}{$T^2$} & \\
\end{tabular}
\end{ruledtabular}
\footnotetext[1]{This phase, the perpendicular conical state, has not been
 experimentally observed so far.}
\footnotetext[2]{The crystal-field effects are much smaller in the NFL region
 than in the ordered phases, see Secs.\ \ref{subsubsec:II.E.4} and
 \ref{subsubsec:III.D.4}. We results listed
 are valid in the universal hydrodynamic regime, which is expected to extend to
 very low temperatures; the true asymptotic low-$T$ behavior is the same as in
 the A-phase, where the size of the $T\to 0$ region is determined by the
 external magnetic field.}
\footnotetext[3]{In the presence of quenched disorder this regime does not
 represent the true asymptotic $T\to 0$ behavior, which is characterized by
 diffusive rather than ballistic electron dynamics, see the table caption and
 Eq.\ (\ref{eq:4.13}).}
\caption{Properties of various ordered phases and the proposed state
 representing the non-Fermi-liquid region. Listed are the number of Goldstone
 modes and their respective dispersion relations, as well as the temperature
 dependence of various observables. The elastic constants ${\tilde c}_{\perp}$,
 $c_z'$, and $c_z$ in the Goldstone modes for the pinned helical/conical phase,
 the A-phase, and the PCS, respectively, are due to crystal-field effects or
 an external magnetic field and are
 small compared to the other elastic constants. The universal hydrodynamic
 regime is the temperature region where crystal-field effects are not important
 and these elastic constants can be neglected. The true asymptotic behavior as
 $T\to 0$ is dominated by the crystal-field effects and is realized only at
 very low temperatures. In the presence of weak disorder, the regime dominated
 by crystal-field effects is bounded below as well as above and does not
 represent the true asymptotic low-temperature regime. See the text for more
 information.}
\label{tab:1}
\end{table*}
In summary, we have given a comprehensive description of all phases in
Dzyaloshinsky-Moriya helical magnets where long-range order has been observed.
These include, the pinned helical phase at weak magnetic fields, the conical
phase at higher magnetic fields, and the A-phase at intermediate magnetic
fields and temperatures close to the critical temperature. We have shown that
the system goes from the conical phase at high magnetic field to the pinned
helical phase at low magnetic field via two distinct phase transitions; a
second-order transition where the orientation of the helix changes smoothly,
followed by a first-order transition where both the orientation and the
polarization of the helix change discontinuously. For the A-phase we have
considered the recent interpretation, based on the observation of a six-fold
symmetry in the neutron scattering signature, as a hexagonal lattice of
skyrmionic line defects. We have also discussed the perpendicular conical
state, which had been discussed earlier as a possible realization of the
A-phase. In addition, we have discussed a proposal for the interpretation of
the non-Fermi-liquid (NFL) region in the disordered phase, which does not
display long-range magnetic order but has many features in common with the
A-phase. For all of these states we have determined the number and nature of
the Goldstone modes, and the temperature dependencies of various observables
that result from the scattering of the conduction electrons by these
excitations. If the state in question can be described by an exact saddle-point
solution of a model Hamiltonian we have provided an explicit calculation; in
other cases we have used symmetry arguments to determine the functional form of
the Goldstone modes. The results are summarized in Table \ref{tab:1}. The
Goldstone modes show unusual anisotropic frequency-momentum relations that
result from helical spin structures in the various states and are reminiscent
of the Goldstone modes in smectic and cholesteric liquid crystals. The specific
heat $C(T)$, the single-particle relaxation rate $1/\tau(T)$ and the thermal
resistivity $\rho_{\text{th}}(T)$, and the electrical resistivity
$\rho_{\text{el}}(T)$, all display temperature dependencies that are distinct
from, and in some cases stronger than, those in a Fermi liquid. The most
remarkable result is a non-Fermi liquid $T^{3/2}$ behavior of the electrical
resistivity in certain temperature regimes in the A-phase and in the NFL
region, which is proposed as an explanation of the observed enigmatic
properties of the DM magnet MnSi.

\acknowledgments

This work was initiated at the Aspen Center for Physics, and supported by the
National Science Foundation under Grant Nos. DMR-09-29966 and DMR-09-01907.

\appendix

\section{The perpendicular conical state}
\label{app:A}

\begin{figure}[t]
\vskip -0mm
\includegraphics[width=7.8cm]{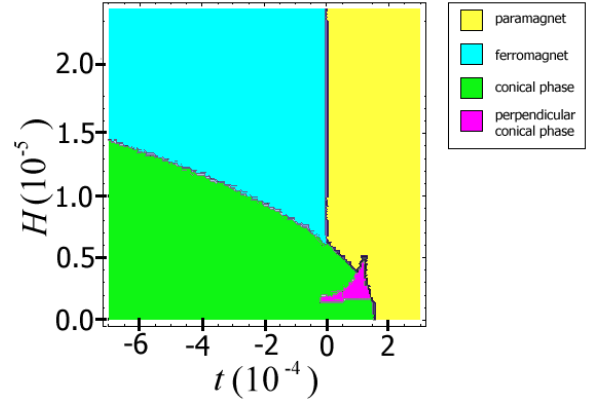}
\caption{Stabilization of the PCS by a large negative value of $w$. Shown are
 the regions of stability for various phases for the action shown in Eq.\
 (\ref{eq:1.3}) with $a=1$, $d=0.025$, $u=0.1$, and $w=-300$.}
\label{fig:8}
\end{figure}

Guided by earlier experimental results that did not reveal the six-fold
symmetry discovered by M{\"u}hlbauer et al.,\cite{Muhlbauer_et_al_2009}
Grigoriev et al.\cite{Grigoriev_et_al_2006a} had proposed a perpendicular
conical state (PCS) to represent the A-phase, i.e., a spin configuration of the
form
\be
{\bm M}({\bm x}) = m_0 {\hat{\bm H}} + m_1 \left[{\hat{\bm e}}_+\, \cos({\bm
q}\cdot{\bm x}) + {\hat{\bm e}}_-\, \sin({\bm q}\cdot{\bm x})\right]
\label{eq:A.1}
\ee
with ${\bm q}$ perpendicular to ${\bm H}$ and ${\hat{\bm e}}_+$, ${\hat{\bm
e}}_-$, and ${\hat{\bm q}}$ forming a {\it dreibein}. This state is not a
saddle point either, and Ref.\ \onlinecite{Muhlbauer_et_al_2009} found it
energetically unfavorable compared to the conical and skyrmion lattice states.
However, it can be stabilized, at least in principle, and we therefore include
it in our discussion. Ref.\ \onlinecite{Grigoriev_et_al_2006a} proposed that
the PCS is stabilized due to a gap in the helimagnon excitation spectrum, which
in turn had been proposed theoretically in Ref.\ \onlinecite{Maleyev_2006}.
Such a gap is at odds with Goldstone's theorem as well as with the calculation
in Ref.\ \onlinecite{Belitz_Kirkpatrick_Rosch_2006a}. A different possibility,
which is consistent with symmetry considerations, is a large negative value of
the parameter $w$ in Eq.\ (\ref{eq:1.3}), which stabilizes the PCS in a
parameter regime close to where the A-phase is observed, see Fig.\ \ref{fig:8}.
While such a large value of $w$ is not realistic, this illustrates that the PCS
can be stabilized by terms in the action that are allowed by symmetry, and it
therefore is useful to determine the properties of this state.

Let us now consider the Goldstone modes in the PCS, using the symmetry
arguments explained in Sec.\ \ref{subsec:III.A}. For definiteness, we assume
that the ${\bm q}$-vector points in the $x$-direction. The ordered state is
still invariant under translations in the $y$ and $z$-directions, so we have
${\text{dim}}(T^3/T^2) = 1$ Goldstone mode. With the magnetic field $H$ in the
$z$-direction, the state is still invariant under rotations of ${\bm q}$ in the
$x$-$y$-plane. The eigenvalue therefore cannot have a $k_y^2$ term. The soft
eigenvalue therefore must have the form
\be
\lambda = \alpha\,k_x^2 + \beta(k_y^4 + k_z^4) + \gamma\,k_z^2,
\label{eq:A.2}
\ee
where $\gamma \propto H^2$ for small $H$. The structure of the dynamics will be
the same as in the aligned conical case, and we thus have a dispersion relation
\be
\omega_{\text{PCS}}({\bm k}) = \sqrt{c_x\,k_x^2 + c_z\,k_z^2 + c_{\perp}(k_y^4
+ k_z^4)},
\label{eq:A.3}
\ee
with $c_z \propto H^2$ and $c_x$ and $c_{\perp}$ constant for $H\to 0$. Here we
neglect pinning effects, which lead to a $k_y^2$-term under the square root at
asymptotically low wave numbers.

For the specific heat contribution we find, from Eqs.\ (\ref{eq:4.1}) and
(\ref{eq:A.3}),
\be
C(T) = {\text{const.}}\times \begin{cases}
           T^{5/2}/(c_x\,c_z)^{1/2}\,c_{\perp}^{1/4} &
                 \text{if $T \ll c_z/\sqrt{c_{\perp}}$} \\
           T^2/\sqrt{c_x\,c_{\perp}} &
                 \text{if $T \gg c_z/\sqrt{c_{\perp}}$}\ .
                             \end{cases}
\label{eq:A.4}
\ee
This result is different from the one for the skyrmionic Goldstone mode in
Sec.\ \ref{subsec:III.C}. Measurements of the specific heat therefore offer a
way to distinguish between these two states short of a direct determination of
the spin structure.

Using Eq.\ (\ref{eq:A.3}) in an obvious generalization of Eqs.\ (\ref{eqs:4.8})
we find for the single-particle relaxation rate and the thermal resistivity
\be
\frac{1}{\tau} \propto \rho_{\text{th}} \propto \begin{cases} T^2 & \text{if $T
\ll
                                                      c_z/\sqrt{c_{\perp}}$}\\
T^{3/2} & \text{if $T \gg c_z/\sqrt{c_{\perp}}$}
                                 \end{cases}\ ,
\label{eq:A.5}
\ee
and for the electrical resistivity
\be
\rho_{\text{el}} \propto \begin{cases} T^3 & \text{if $T \ll
     c_z/\sqrt{c_{\perp}}$}\\
T^{5/2} & \text{if $T \gg c_z/\sqrt{c_{\perp}}$}
                                 \end{cases}\ .
\label{eq:A.6}
\ee

\section{Gaussian fluctuations in the aligned conical phase}
\label{app:B}

Here we derive the Gaussian fluctuation action in the conical phase, Eq.\
(\ref{eq:3.7}). The equation of state for this phase, Eqs.\ (\ref{eqs:2.6},
\ref{eqs:2.7}), can be written
\be
t - aq^2 + u(m_0^2 + m_1^2) = 0,
\label{eq:B.1}
\ee
with $q$ and $m_0$ given by Eqs.\ (\ref{eq:2.6b}) and (\ref{eq:2.6c}),
respectively. Inserting Eq.\ (\ref{eq:3.10}) into the action, Eqs.\
(\ref{eq:1.3}, \ref{eq:1.4a}), dropping the amplitude fluctuations $\delta m_0$
and $\delta m_1$, and expanding to bilinear order in the phases yields Eq.\
(\ref{eq:3.7}) with
\begin{widetext}
\be
\gamma({\bm k}) = \left(\begin{array}{ccccc}
   {\hat m}_1^2\,{\hat k}^2 & -i {\hat m}_1^2 {\hat k}_y & i {\hat m}_1^2 {\hat
                                                             k_x} & 0 & 0 \\
   i {\hat m}_1^2 {\hat k}_y & {\hat m}_1^2 (1 + {\hat m}_0^2 +
      \frac{1}{2}\,{\hat k}^2) & -i {\hat m}_1^2 {\hat k}_z & 0 &
      {\hat m}_0^2 {\hat m}_1^2 \\
   -i {\hat m}_1^2 {\hat k_x} & i {\hat m}_1^2 {\hat k}_z
                              & {\hat m}_1^2 (1 + {\hat m}_0^2 +
      \frac{1}{2}\,{\hat k}^2) & {\hat m}_0^2 {\hat m}_1^2 & 0 \\
   0 & 0 & {\hat m}_0^2 {\hat m}_1^2 & {\hat m}_0^2 (1 + {\hat m}_1^2 + {\hat
      k}^2) & 2i {\hat m}_0^2 {\hat k}_z \\
   0 & {\hat m}_0^2 {\hat m}_1^2 & 0 & -2i {\hat m}_0^2 {\hat k}_z &
      {\hat m}_0^2 (1 + {\hat m}_1^2 + {\hat k}^2)
      \end{array} \right).
\label{eq:B.2}
\ee
A determination of the eigenvalues yields Eqs.\ (\ref{eqs:3.8}, \ref{eqs:3.9}).

\section{Gaussian fluctuations in the pinned helical phase}
\label{app:C}

Here we derive the Gaussian fluctuation action in the pinned helical phase,
Eq.\ (\ref{eq:3.15}). Dropping the amplitude fluctuations, and expanding to
linear order in the phase fluctuations, Eq.\ (\ref{eq:3.14}) becomes
\bse
\label{eqs:C.1}
\be
{\bm M}({\bm x}) = {\bm M}_{\text{sp}}({\bm x}) + \delta{\bm M}({\bm x}),
\label{eq:C.1a}
\ee
with
\be
\delta{\bm M}({\bm x}) = m_1\bigl[(-{\hat{\bm e}}_{-}\,\sin({\bm q}\cdot{\bm
x}) + {\hat{\bm e}}_{+}\,\cos({\bm q}\cdot{\bm x}))\,\varphi_0({\bm x}) +
{\hat{\bm q}}\,\sin({\bm q}\cdot{\bm x})\,\varphi_{-}({\bm x}) + {\hat{\bm
q}}\,\cos({\bm q}\cdot{\bm x})\,\varphi_{+}({\bm x}) + O(\varphi_0^2)\bigr].
\label{eq:C.1b}
\ee
\ese
${\bm M}_{\text{sp}}$ is an exact saddle point of the action, Eqs.\
(\ref{eq:1.2}) - (\ref{eqs:1.5}) with $b_1 = v = 0$, so the terms linear in
$\delta{\bm M}$ vanish. We now consider terms bilinear in $\varphi_{0,+,-}$ and
first concentrate on the gradient-free terms. Neglecting rapidly fluctuating
Fourier components proportional to $e^{in{\bm q}\cdot{\bm x}}$ with $n \geq 2$
one finds
\bea
{\cal A}^{(2)}[\delta{\bm M}]/m_1^2 &=& \frac{1}{2} \int d{\bm x}\
\varphi_0({\bm x})\,\left[t + aq^2 - cq + um_1^2 + \frac{1}{2}\,b \sum_{i=1}^3
q_i^2\left[ ({\hat e}_+^i)^2 + ({\hat e}_-^i)^2\right]\right]\,\varphi_0({\bm
x})
\nonumber\\
&&+ \frac{1}{4} \int d{\bm x}\ \sum_{\alpha=\pm} \varphi_{\alpha}({\bm
x})\,\left[t + aq^2 + u m_1^2 + b\,q^2\sum_{i=1}^3 {\hat q}_i^4
\right]\,\varphi_{\alpha}({\bm x})
\nonumber\\
&&-\frac{1}{2}\,b\,q^2 \int d{\bm x}\ \sum_{\alpha=\pm} \sum_{i=1}^3 {\hat
e}_{\alpha}^i {\hat q}_i^3\,\varphi_0({\bm x})\,\varphi_{\alpha}({\bm x}) +
\text{(gradient terms)}
\label{eq:C.2}
\eea
\end{widetext}
An explicit calculation shows that
\bse
\label{eqs:C.3}
\bea
\sum_{i=1}^{3} {\hat q}_i^2 \left[({\hat e}_+^i)^2 + ({\hat e}_-^i)^2\right]
&=& 1 - f({\hat{\bm q}}),
\label{eq:C.3a}\\
\sum_{i=1}^{3} {\hat q}_i^4 = f({\hat{\bm q}}),
\label{eq:C.3b}
\eea
with
\be
f({\hat{\bm q}}) = \beta_1^4 + \beta_2^4 + \beta_3^4,
\label{eq:C.3c}
\ee
where $\beta_{1,2,3}$ are the direction cosines of ${\hat{\bm q}}$.
Furthermore,
\be
\sum_{i=1}^{3} {\hat e}_{\alpha}^i\, ({\hat q}_i)^3 = 0
\label{eq:C.3d}
\ee
\ese
for $\alpha = +,-$ and for both ${\hat{\bm q}} = (1,1,1)/\sqrt{3}$ and
${\hat{\bm q}} = (1,0,0)$. Using the equation of state, Eqs.\ (\ref{eq:3.13b},
\ref{eq:3.13c}), we see that the $\varphi_0$-$\varphi_0$ vertex and the
$\varphi_0$-$\varphi_{\alpha}$ vertices all vanish. At zero wave number we thus
have one zero eigenvalue that corresponds to one Goldstone mode, in agreement
with the expectation from Sec.\ \ref{subsubsec:III.B.1}.

We next calculate the gradient-squared terms. Geometric identities similar to
those expressed in Eqs.\ (\ref{eqs:C.3}) allow to determine the vertices. One
finds a Gaussian action of the form given by Eq. (\ref{eq:3.15}). For $b>0$,
the matrix $\Gamma$ is given by
\begin{widetext}
\be \Gamma({\bm k})
= \left(\begin{array}{ccc}
                              a\,{\bm k}^2 + \frac{1}{2}\,b\,{\bm k}_{\perp}^2 &
                                    -i\,\frac{1}{2}\,c\,k_y                  &
                                    -i\,\frac{1}{2}\,c\,k_x                  \\
                              i\frac{1}{2}\,c\,k_y                         &
                                   \frac{1}{2}\,c\,q + \frac{1}{2}\,a\,{\bm k}^2
                                              + \frac{1}{2}\,b\,k_z^2        &
                                   i(a+b)qk_z                             \\
                              i\,\frac{1}{2}\,c\,k_x                         &
                                   -i(a+b)\,q\,k_z                            &
                                   \frac{1}{2}\,c\,q + \frac{1}{2}\,a\,{\bm k}^2
                                               + \frac{1}{2}\,b\,k_z^2
                         \end{array}\right)
\qquad \begin{array}{ccc}  \\  \\  (b>0)\quad . \end{array}
\label{eq:C.4}
\ee
For $b<0$, the matrix $\Gamma$ takes the form
\bea
\Gamma_{00}({\bm k}) &=& (a + b/3)\,k^2,
\nonumber\\
\Gamma_{0+}({\bm k}) &=& \Gamma_{+0}({\bm k})^* = i\left(\frac{1}{2}\,c +
\frac{1}{3}\,b\,q\right)\frac{1}{\sqrt{2}}\,(k_{||} - 3k_z) +
\frac{1}{2\sqrt{6}}\,b\,(k_x^2 - k_y^2),
\nonumber\\
\Gamma_{0-}({\bm k}) &=& \Gamma_{-0}({\bm k})^* = i\left(\frac{1}{2}\,c +
\frac{1}{3}\,b\,q\right)\frac{1}{\sqrt{2}}\,(k_x - k_y) -
\frac{1}{2\sqrt{6}}\,b\,(k_x^2 + k_y^2 - 2k_z^2),
\nonumber\\
\Gamma_{++}({\bm k}) &=&  \Gamma_{--}({\bm k}) = \frac{1}{2}\,c\,q +
\frac{1}{2}\,(a + b/3)k^2,
\nonumber\\
\Gamma_{+-}({\bm k}) &=& \Gamma_{-+}({\bm k})^* = i(a+b/3) q\,k_{||} \hskip
160pt (b<0)\quad .
\label{eq:C.5}
\eea
\end{widetext}
Calculating the eigenvalues one finds Eq.\ (\ref{eq:3.16}).

\section{Relaxation rates in ferromagnets}
\label{app:D}

In this appendix we sketch how the expressions for the single-particle and
transport relaxation rates given in Sec.\ \ref{subsec:IV.B} change in the
ferromagnetic limit. A more thorough discussion of this topic will be given
elsewhere.\cite{us_tbp}

There are three major changes compared to Eqs.\ (\ref{eq:4.4}) and
(\ref{eq:4.7}) that take place in the ferromagnetic limit. Firstly, there is no
intra-Stoner-band scattering. This is because in a ferromagnet, the Goldstone
mode is entirely transverse with respect to the direction of the magnetization,
whereas in a helimagnet this is not the case.\cite{transverse_footnote} As a
result, with ${\bm k}$ in Eq.\ (\ref{eq:4.4}) on the 1-Fermi surface, the
resonance frequency inside the delta function under the integral will be
$\omega_2({\bm p})$. Secondly, the dimensionless vertex $\tilde\gamma$, Eq.\
(\ref{eq:4.6b}), changes to unity. Technically, this is seen most easily within
the formalism of Ref.\ \onlinecite{Kirkpatrick_Belitz_Saha_2008a}. Physically,
it reflects the fact that for intra-Stoner-band scattering the magnetic
fluctuations couple to the quasi-particle density, and thus physically act akin
to a chemical potential, only gradients of which contribute to scattering. For
inter-Stoner-band scattering, on the other hand, they physically act akin to an
external magnetic field, and thus they couple without gradients. Thirdly, the
form of the susceptibility $\chi$, Eq.\ (\ref{eq:4.6c}), changes. It now
reflects the ferromagnetic magnons, and for power counting purposes it can be
adequately represented by
\bse
\label{eqs:D.1}
\be
\chi''({\bm k},u) = \frac{1}{\NF\lambda}\, \delta(u - \omega_{\text{FM}}({\bm
k})),
\label{eq:D.1a}
\ee
with
\be
\omega_{\text{FM}}({\bm k}) = {\rm const.}\times\lambda k^2/\kF^2
\label{eq:D.1b}
\ee
\ese
the ferromagnon frequency. Putting all of this together, and neglecting
numerical prefactors, we obtain for the single-particle scattering rate due to
ferromagnetic magnons
\bse
\label{eqs:D.2}
\be
\frac{1}{\tau({\bm k})} = \frac{\lambda}{\NF}\,\frac{1}{V}\sum_{\bm p}
\frac{1}{\sinh(\omega_{\text{FM}}({\bm p})/T)}\ \delta(\omega_2({\bm k} + {\bm
p}) - \omega_{\text{FM}}({\bm p})),
\label{eq:D.2a}
\ee
and for the electrical transport rate
\bea
\frac{1}{\tau_{\text{el}}^{\text{tr}}} &=& \frac{\lambda}{\NF^2} \frac{1}{V^2}
\sum_{{\bm p},{\bm k}} \frac{p^2/\kF^2}{\sinh(\omega_{\text{FM}}({\bm p})/T)}\
\delta(\omega_1({\bm k}))
\nonumber\\
&&\hskip 40pt \times \delta(\omega_2({\bm k}+{\bm p}) - \omega_{\text{FM}}({\bm
p}))\ .
\label{eq:D.2b}
\eea
\ese

Performing the integrals leads to the following results. For the
single-particle relaxation rate on the Fermi surface, or the thermal
resistivity, one finds
\be
\frac{1}{\tau} \propto \rho_{\text{th}} \propto \begin{cases} T\,
e^{-\lambda^3/\epsilonF^2 T} &
\text{if $T\ll\lambda^3/\epsilonF^2$}\\
   T\,\ln(T\epsilonF^2/\lambda^3) & \text{if $\lambda^3/\epsilonF^2 \ll T \ll
   \lambda$}\\
   T\,\ln(\epsilonF^2/\lambda^2) & \text{if $T \gg \lambda$}\ ,
   \end{cases}
\label{eq:D.3}
\ee
where we have omitted numerical prefactors as well as less-leading terms. For
the electrical resistivity, the corresponding result is
\be
\rho_{\text{el}} \propto \begin{cases} T\,(\lambda^2\epsilonF^2)\,
   e^{-\lambda^3/\epsilonF^2 T} & \text{if $T\ll\lambda^3/\epsilonF^2$}\\
   T^2/\lambda & \text{if $\lambda^3/\epsilonF^2 \ll T \ll
   \lambda$}\\
   T & \text{if $T \gg \lambda$}\ ,
   \end{cases}
\label{eq:D.4}
\ee
We provide some brief comments regarding these results, a more complete
discussion will be given elsewhere.\cite{us_tbp} (1) The energy scales
$\lambda^3/\epsilonF^2$ and $\lambda$ that lead to the three temperature
regimes shown above emerge within the bare (Stoner-level) theory. It is not
obvious how these scales become renormalized, and one therefore has to be
careful when making quantitative comparisons with experiment. In particular,
within the bare theory the regime $T \gg \lambda$ is not realizable since it
implies $T > \Tc$; this may change within a fully renormalized theory. (2) The
second result in Eq.\ (\ref{eq:D.4}), $1/\tau_{\text{el}}^{\text{tr}} \propto
T^2/\lambda$, reproduces the result of Ueda and Moriya.\cite{Ueda_Moriya_1975}
We note, however, that this is {\em not} the true asymptotic low-temperature
result; it is valid only in a pre-asymptotic temperature window. (3) The true
asymptotic behavior for $T\to 0$ is exponential, rather than power-law, in $T$.
This is due to the Stoner splitting of the conduction electrons; this effect
was neglected in Ref.\ \onlinecite{Ueda_Moriya_1975}. In helimagnets the
situation is qualitatively different, see Sec.\ \ref{subsec:IV.B}.


\begin{thebibliography}{58}
\bibitem[{cor()}]{corrections_footnote}
\bibinfo{note}{More precisely, the criterion is $b < b_1/2 - 3b_1^2 q/c$. The
  reason for this correction is the $(\delta\pi)^2$ term in the free energy,
  Eq.\ (\ref{eq:2.10a}). We neglect this and other corrections that are of
  higher order in $\gso$. For instance, there are higher-order corrections to
  $\Hcone$ as given by Eq.\ (\ref{eq:2.13b}).}
\bibitem[{rat()}]{rates_footnote}
\bibinfo{note}{It is remarkable that in this regime the difference between the
  single-particle relaxation rate and the electrical transport rate that is so
  characteristic of the Boltzmann theory of transport does not exist. From a
  scaling point of view one might speculate that this difference is not generic
  and will not be present in a strong-coupling theory that would be
  characterized by a single universal relaxation rate. Whether this is true,
  and if so, in what limit the Boltzmann result emerges, deserves an
  investigation.}
\end{thebibliography}

\end{document}